\documentclass[12pt]{article}
\usepackage[table]{xcolor}
\definecolor{lightgray}{gray}{0.9}
\usepackage[onehalfspacing]{setspace}
\usepackage[linesnumbered,ruled,vlined]{algorithm2e}
\usepackage{amsmath,amssymb,amsfonts,amsthm}
\usepackage[final]{graphicx}
\usepackage{pgfpages}
\usepackage{rotating}
\usepackage{pdflscape}
\usepackage{tablefootnote}
\usepackage{pdfpages}
\usepackage{float}
\usepackage{bm}
\usepackage{bbm}
\usepackage{multirow}
\usepackage[normalem]{ulem}
\usepackage{multicol}
\usepackage{booktabs}

\usepackage{lscape}
\usepackage{enumitem}
\usepackage[font=footnotesize]{caption}
\usepackage{physics}
\usepackage{siunitx}
\usepackage{geometry}
\usepackage{authblk}
\usepackage{tikzsymbols}
\usepackage{fontawesome}
\usepackage{pifont}
\usepackage{scrextend}
\usepackage{subcaption}
\usepackage{tikz}
\usepackage{natbib}
\usepackage[breaklinks]{hyperref}
\hypersetup{
    colorlinks = true,
    linkcolor=blue,
    citecolor=blue,
    urlcolor=cyan
    }
\newcommand{\mycite}[2]{\hyperlink{cite.#1}{\textcolor{blue}{#2}}}
\newcommand{\AFV}{\mycite{AF15}{AFV}\@}
\usepackage{titlesec}
\usepackage{tikz}
\usetikzlibrary{arrows.meta}
\usepackage{pgfplots}
\pgfplotsset{width=0.75\linewidth,compat=1.9}
\newtheorem{theorem}{Theorem}

\newtheorem{definition}[theorem]{Definition}

\begin{document}

\title{Dynamic Programming on a Quantum Annealer:\\ Solving the RBC Model\thanks{We thank participants at numerous seminars and conferences for useful comments and suggestions. During part of the project, Isaiah Hull was affiliated with and had a financial interest in CogniFrame, Inc., which is a partner of D-Wave Systems, the producer of the quantum annealers used in the paper. The authors did not receive any support or compute time from D-Wave through this partnership. All access to D-Wave's quantum annealers was provided through a standard developer agreement.}}

\author{Jes\'{u}s Fern\'{a}ndez-Villaverde \\
{University of Pennsylvania, NBER, and CEPR} \and Isaiah Hull\\
{\hspace{0.15in} Department of Finance, BI Norwegian Business School, \\ and CogniFrame Inc.}}

\date{\today}

\maketitle

\vspace{-0.2in}

\begin{abstract}
We introduce a novel approach to solving dynamic programming problems, such as those in many economic models, on a quantum annealer, a specialized device that performs combinatorial optimization. Quantum annealers attempt to solve an NP-hard problem by starting in a quantum superposition of \textit{all} states and generating candidate \textit{global} solutions in milliseconds, irrespective of problem size. Using existing quantum hardware, we achieve an order-of-magnitude speed-up in solving the real business cycle model over benchmarks in the literature. We also provide a detailed introduction to quantum annealing and discuss its potential use for more challenging economic problems.

\vspace{5mm}


\vspace{4mm}


\end{abstract}

\newpage
 
\section{Introduction}\label{sec:Introduction}

We introduce a new approach to solving dynamic programming problems, such as those that appear in many economic models, on a quantum annealer (QA). This specialized quantum device performs combinatorial optimization using a physical process. A QA embeds the problem's parameters into a quantum system that evolves to find its lowest energy configuration. This is equivalent to finding the values of state variables that globally minimize a loss function \citep{FGGS00}. QAs attempt to solve a problem that is NP-hard for a classical computer by starting in a quantum superposition of all states and returning candidate solutions in milliseconds, irrespective of the problem size \citep{VCML18}.\footnote{We use the term ``classical'' and ``classically'' to indicate that something is not quantum.}

Our paper makes four contributions:

\begin{enumerate}

\item The development of a new solution method for solving dynamic programming problems on quantum hardware.

\item The implementation and execution of the solution method on existing quantum hardware, rather than on a classical simulator.

\item The novel use of state-of-the-art quantum annealing techniques, including reverse and inhomogeneous annealing.

\item Showing the broad applicability of our solution method to iterative problems that could not otherwise be solved on QAs without hybridizing the problem into classical and quantum components.

\end{enumerate}

More concretely, we tackle the limitations of QAs, which are not designed to solve the dynamic programming problems at the core of many economic models. In particular, QAs do not naturally allow for iteration over time or across multiple objective functions and suffer from the quantum-to-classical bottleneck, which severely limits how much classical information can be read out as the problem's solution. Our approach overcomes these limitations and can be used to recover policy and value functions for problems in macroeconomics, industrial organization, game theory, and labor economics.

To evaluate our approach, we solve the real business cycle (RBC) model on a QA and compare its performance to the benchmark results in \cite{AF15} (hereafter, \AFV{}). Solving the RBC model also allows us to demonstrate how to formulate a well-known economic model in a way that can be solved on a QA. Even with the limitations of existing quantum technology, we can solve the RBC model on a QA in 3\% of the computation time of the VFI solution using \texttt{C++} as in \AFV{} or 0.66\% of the computation time of the combinatorial algorithm that we propose for the QA but run on a classical computer. Thus, we demonstrate the enormous potential of quantum hardware for economists.

Our approach differs from the existing literature on quantum dynamic programming in three key aspects. First, we use a QA rather than a universal quantum computer (UQC).\footnote{Section \ref{sec:QuantumAnnealing} will explain in more detail the difference between QAs and UQCs. Suffice it to say here that UQCs are more challenging to construct than QAs, which is why their development has lagged.} While UQCs allow for the proof of reductions in computational complexity, they are not sufficiently mature to implement dynamic programming algorithms that offer a quantum speed-up for non-trivial problems. Prior work on dynamic programming on UQCs has focused, instead, on the theoretical demonstration of quantum speed-ups and on proof-of-principle demonstrations for trivial problems.\footnote{Even on UQCs, little theoretical progress had been made before \citet{ABI+19}, who used the unstructured quantum search algorithm in \citet{Gro96}, coupled with the computation of a partial dynamic programming table, to achieve a quadratic speed-up for problems that involve the selection of a subset of elements.} In contrast, our focus on QAs allows us to solve a standard RBC model on the current vintage of quantum hardware rather than simulating not-yet-developed quantum hardware on a classical computer.

Second, we focus on problems of interest to economists and explore what is achievable given already existing machines. Work on dynamic programming using quantum devices has centered almost exclusively on problems in physics and computer science that have naturally parsimonious solutions. In contrast, economists are typically concerned with recovering policy or value functions that live in high-dimensional hypercubes. This is a non-trivial difference since the quantum-to-classical bottleneck severely limits how much data can be read out of a quantum device. For instance, a QA with $N$ qubits will start in an exponentially large quantum superposition state of dimension $2^{N}$, but will eventually collapse into a classical state that contains only $N$ classical bits of information. Thus, finding a way to encode the solution in $N$ bits is a fundamental challenge for dynamic programming problems in economics, while it is usually not a primary concern in physics and computer science. In part, we overcome this encoding problem by using the parametric dynamic programming method introduced in \citet{BRG+00} to reduce the space complexity of the solution.

Third, we show how to construct iterative algorithms that can be implemented on QAs, which no other work has accomplished to the best of our knowledge. A fundamental challenge of dynamic programming is that solution methods often require iteration, sometimes over alternating objective functions. QAs are not programmable in the way that classical computers or UQCs are programmable. That is, they can only embed a problem instance and then perform annealing with a limited set of parameters. Our approach addresses the problem of how to implement iterations by making use of two state-of-the-art features of QAs: reverse and inhomogeneous annealing. These two features widen the applicability of our work to any iterative algorithm, whether or not it is related to dynamic programming.

The rest of the paper is structured as follows. Section \ref{sec:QuantumAnnealing} gives an overview of quantum annealing. Section \ref{sec:Model} introduces the benchmark RBC model we use as a testbed. In Section \ref{sec:Methods}, we explain how to translate the model into a form that is solvable on a QA. Section \ref{sec:Results} introduces hybrid and quantum algorithms for solving the model and compares the results to the benchmarks in \AFV{}. Section \ref{sec:Discussion} concludes with a discussion of the prospects of quantum annealing in economics. An Online Appendix adds further details. 

\section{Quantum Annealing}\label{sec:QuantumAnnealing}

There are two primary models of quantum computing: universal quantum computing and quantum annealing. Universal quantum computers (UQCs) employ the ``gate-and-circuit'' model, in which quantum operations called gates are applied to quantum bits (qubits) to enable arbitrary computations. These gates must adhere to the principles of quantum physics, since they are performing an operation on a quantum system. Quantum circuits consist of gate sequences that implement a subroutine within a quantum system.

UQCs offer theoretical reductions in time and space complexity for various algorithms used in computational economics and econometrics \citep{HSDW20}. For instance, \citet{ABI+19} and \citet{GKM+21} prove theoretical quantum speed-ups for several classes of dynamic programming problem, including the traveling salesperson and vertex ordering problems. Experimental evidence also demonstrates that UQCs are capable of achieving a speed-up over classical computers for certain problem classes \citep{AAB+19}.\footnote{\cite{AAB+19} initially claimed to achieve ``quantum supremacy,'' as defined in \cite{Pre12}, by performing a computation in 200 seconds that would arguably take 10,000 years on the world's fastest classical supercomputer; however, this claim has since been disputed and appears to have understated the performance of classical supercomputers.} Unfortunately, UCQs are technically demanding to construct, and their development has lagged behind specialized quantum devices, such as QAs.

In contrast, QAs do not offer theoretically provable reductions in computational complexity and cannot be programmed to execute arbitrary algorithms. Instead, QAs can only perform combinatorial optimization. This seemingly severe limitation has, however, enabled considerably faster experimental progress. At present, QAs have roughly 50 times as many qubits as UQCs.\footnote{The largest UQCs are currently produced by IBM. Its \textit{Eagle r1} quantum processor has 127 qubits. IBM has also recently introduced an exploratory \textit{Osprey r1} processor with 433 qubits. In contrast, \textit{Advantage}, the QA produced by D-Wave Systems that we use for our computations in this paper, has 5616 qubits.}

In the remainder of this section, we introduce the concept of quantum annealing, discuss the problems it can solve, explain its limitations, and identify problems in economics where it might be applied. We start with a brief review of combinatorial optimization.

\subsection{Combinatorial Optimization}\label{sec:combinatorial_optimization}

Quantum annealing is a heuristic method for solving combinatorial optimization problems that may provide a substantial computational advantage over classical solution methods for specific cases \citep{ZBT+15}. For our discussion, we adopt a slightly modified version of the definition of a combinatorial optimization problem given in \citet{VCML18}.

\begin{definition}[Combinatorial Optimization Problem] 
Let $E$ be a finite set with cardinality $|E| = n$, $P_{E}$ the power set of $E$ (hence $|P_{E}| = 2^{n}$), and $C$ a loss function where $C:P_{E} \rightarrow \mathbb{R}$. The general setup of a combinatorial optimization problem is to find an element $\mathcal{P} \in P_{E}$ such that $C(P) = min_{\mathcal{P}_{i} \in P_{E}} \{ C(\mathcal{P}_{i}) \}$.
\end{definition}

In game theory and mechanism design, problems involving the allocation of discrete objects often have a natural combinatorial form that fits the definition above and is conducive to using a QA. Problems that involve financial networks also align with this definition and can be solved on a QA, as shown in proof-of-principle exercises by \citet{OML19b,OML19a}.

Our focus will be on problems in economics that do not naturally fit this definition but are commonly reformulated as combinatorial problems to facilitate grid search or value function iteration. As the first example, consider a firm that minimizes a cost function, $C$, by choosing inputs, $L$ and $K$, given technology $y^{*} = Y(L,K)$. If the problem does not permit an analytical solution, a common strategy is to search for the minimum of $C$ over the Cartesian product of discrete grids, where $K \in \{ K_{0}, K_{1}, ..., K_{\bar{K}-1} \}$ and $L \in \{ L_{0}, L_{1}, ..., L_{\bar{L}-1} \}$. A solution to the problem is a pair of functions, $K^{*}(w,r,y)$ and $L^{*}(w,r,y)$, which yield optimal $K$ and $L$ choices given factor prices $(w,r)$ and the target level of output $y$.

As a second example, consider the problem of an infinitely lived household that maximizes utility from consumption, subject to a budget constraint. This problem is often written as a dynamic program and solved using value function iteration. The solution is a discrete look-up table approximation of the optimal value function or policy rule for capital and consumption.

While these two examples can, in principle, be solved on a QA once we have reformulated them as combinatorial problems, the limitations of quantum computers will make this reformulation challenging. The primary difficulty, which we discuss in Subsection \ref{ssec:limitations}, is the quantum-to-classical bottleneck, which limits the amount of information that can be read out of a quantum computer, including a QA. The computational complexity of such problems also differs from that of more standard combinatorial optimization problems, which we discuss in Subsection \ref{ssec:computational_complexity}.

\subsection{Adiabatic Quantum Computing}\label{ssec:aqc}

Quantum annealing can be described as a \textit{heuristic} or \textit{finite temperature} implementation of adiabatic quantum computing \citep{VCML18, SSSV14}, and its value as an optimization algorithm is justified through the quantum adiabatic theorem \citep{Mes58, AR04}. To explain these ideas, let us introduce the idealized concept of adiabatic quantum computing, which we will treat as the best-case scenario for quantum annealing that is unlikely to be achieved in practice.

\citet{AR04} provide an informal definition of the \textit{quantum adiabatic theorem}. A Hamiltonian expresses the total level of energy in a quantum system:

\begin{quote}
\textit{... if we take a quantum system whose Hamiltonian slowly changes from $\mathcal{H}_{1}$ to $\mathcal{H}_{2}$, then, under certain conditions on $\mathcal{H}_{1}$ and $\mathcal{H}_{2}$, the ground (lowest energy) state of $\mathcal{H}_{1}$ gets transformed to the ground state of $\mathcal{H}_{2}$.}
\end{quote}

\citet{FGGS00} showed that the previous fact can be used to construct an alternative model of quantum computing that exploits adiabatic evolution to identify global minima.

More concretely, one builds a quantum state that interpolates between two Hamiltonians: 1) a trivial Hamiltonian, $\mathcal{H}_{0}$; and 2) the problem Hamiltonian, $\mathcal{H}_{p}$. The system is initialized as $\mathcal{H}_{0}$, but transitions over time to $\mathcal{H}_{p}$, following the adiabatic evolution:
\begin{equation*}
    \mathcal{H}(t) = \left( 1 - \frac{t}{T} \right) \mathcal{H}_{0} + \frac{t}{T} \mathcal{H}_{p}.
\end{equation*}
The initial Hamiltonian, $\mathcal{H}_{0}$, is specified to be trivial, such that we can identify the lowest level of energy (the ground state) analytically. In comparison, $\mathcal{H}_{p}$ encodes a minimization problem of interest, where the energy level in the system corresponds to the loss associated with a state. Thus, $\mathcal{H}_{0}$ and $\mathcal{H}_{p}$ can be interpreted as loss functions with ground states that correspond to global minima.

Returning to the description of the quantum adiabatic theorem, evolving from $\mathcal{H}_{0}$ to $\mathcal{H}_{p}$ sufficiently slowly will ensure that the quantum system remains in the global minimum (ground state) in all periods. Thus, in principle, the quantum adiabatic theorem tells us how to find the global minimum of a combinatorial optimization problem.

Adiabatic quantum computing also allows us to compute a ``speed limit'' for the transition between $\mathcal{H}_{0}$ and $\mathcal{H}_{p}$ to ensure the quantum system remains in its ground state. Unfortunately, for many problem types of size $N$, the speed limit takes the form $T = \mathcal{O}[exp(\gamma N^\nu)]$, where $\gamma$ and $\nu$ are positive parameters \citep{BFK+13, Luc14}.

Hence, adiabatic quantum computing may require exponential time to transition between $\mathcal{H}_{0}$ and $\mathcal{H}_{p}$ while remaining in the ground state. Consequently, it is unlikely that QAs will solve hard global optimization problems in polynomial time. However, $\gamma$ and $\nu$ could be considerably smaller than their corresponding values for the classical problem, enabling us to solve exponentially hard problems with larger input sizes. We will revisit this idea in Subsection \ref{ssec:computational_complexity}.

\subsection{Problem Encoding}\label{ssec:encoding}

Quantum annealing involves the evolution of a trivial Hamiltonian, $\mathcal{H}_{0}$, configured in its ground state, into one that encodes the problem of interest, $\mathcal{H}_{p}$. If the transition happens adiabatically (very slowly), $\mathcal{H}_{p}$ will also be in its ground state. We can then measure the quantum system and recover the global minimum.

QAs do not require us to specify $\mathcal{H}_{0}$, since any trivial Hamiltonian with a known ground state (global minimum) will work.\footnote{In practice, an equal superposition of all states is typically used. Additionally, ``reverse'' annealing allows us to specify an initial classical state, which can be used to refine solutions by performing a local search.} However, one must specify $\mathcal{H}_{p}$, a non-trivial task even for simple problems. This process is referred to as ``problem encoding,'' since it encodes the information about a minimization problem in the state of a quantum system. Next, we will discuss how such encodings are constructed, starting with the types of problems we can feasibly encode.

\subsubsection{Conversion to a binary quadratic model}\label{sssec:ising_and_qubo}

Subsection \ref{sec:combinatorial_optimization} highlighted that an optimization problem must be combinatorial to be solvable on a QA. This is because QAs must encode the structure of a minimization problem as a binary quadratic model (BQM). This can be done using the Ising or the quadratic unconstrained binary optimization (QUBO) model. The Ising model allows us to introduce the main concepts transparently and has an intuitive interpretation. The QUBO will be more convenient and, thus, our preferred choice for the rest of the paper. The QUBO model also abstracts from any physical system and, thus, does not require knowledge of physics.

\paragraph{The Ising model.} The Ising model describes a system of atomic spins. For our purposes, it will suffice to understand that each spin is a physical system that can be configured in one of two states, which we will represent as +1 and -1. We can think of a spin as corresponding to a terminal qubit state in a QA.

We denote a vector of $N$ spins as $s = \{ s_{0},s_{1},...,s_{N-1} \}$. Each $s_{i}$ has a bias, $h_{i}$, and each pair of spins, $(s_{i},s_{j})$, has a coupling, $J_{i,j}$, where $h_{i},J_{i,j} \in \mathbb{R}$. The spins and couplings can be used to define an Ising model:
\begin{equation*}
    \mathcal{H}(s) = \sum_{i=0}^{N-1} h_{i} s_{i} + \sum_{i=0}^{N-1} \sum_{j=0}^{N-1} J_{i,j} s_{i} s_{j}.
\label{eqn:ising_model}
\end{equation*}

Recall that $\mathcal{H}$ encodes a loss function as the total energy level in a physical system. Thus, a terminal $s$ vector that implies a higher $\mathcal{H}$ also implies a higher loss. The biases and couplings are features of our optimization problem, which we encode in the system. The annealing process identifies an optimal configuration of spins to minimize the energy level.

\begin{table}[htbp!]
	\centering
	\begin{tabular}{|l|l|l|l|}
	    \hline
		& $s_{0}$ & $s_{1}$ & $\mathcal{H}(s)$ \\ \hline
		1 & -1 & -1 & -1.0 \\
		2 & -1 & 1 & 0.0 \\
		3 & 1 & -1 & 1.6 \\
		4 & 1 & 1 & -0.6 \\
		\hline
	\end{tabular}
\caption{Spin configurations and energy levels.}
\label{tab:ising_model}
\end{table}

For concreteness, consider a problem where $N=2$, $h = \{0.5, -0.3\}$, and $J = \{-0.8\}$. Table \ref{tab:ising_model} enumerates all possible terminal $s$ vectors and computes the total energy (loss) in the system. Combination 1, where $s = \{-1, -1\}$, is a global minimum for this problem, since $\mathcal{H}(-1,-1)$ is lower than for any other combination of spins. While this is trivial to prove for the $N=2$ case, the classical time complexity of enumerating all possible combinations for an arbitrary $s$ of length $N$ is $\mathcal{O}(2^{N})$.

\begin{figure}[htbp!]
	\begin{center}
	\begin{tikzpicture}

	\begin{scope}[every node/.style={black,circle,thick,draw}]
    		\node (0) at (-2,0) {$s_{0}$};
    		\node (1) at (0,2) {$s_{1}$};
    		\node (2) at (2,0) {$s_{2}$};
    		\node (3) at (0,-2) {$s_{3}$};
	\end{scope}

	\begin{scope}[>={Stealth[black]},
		every edge/.style={draw=black,very thick}]
		
	\path [-] (0) edge (1);
	\path [-] (0) edge (3);
	\path [-] (1) edge (2);
	\path [-] (1) edge (3);
	\end{scope}

	\end{tikzpicture}

	\caption{A graphical representation of a binary quadratic model for the $N=4$ case and with couplings between the pairs: $\{ (s_{0},s_{1})$,$(s_{0},s_{3})$, $(s_{1},s_{2})$,$(s_{1},s_{3})\}$. }
\label{fig:graphical_representation}
\end{center}
\end{figure}
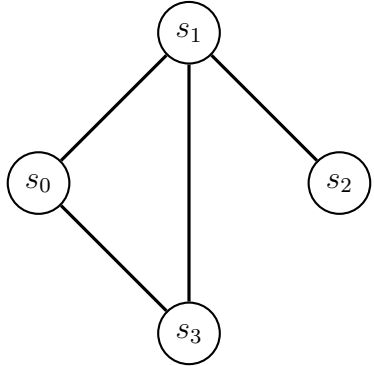

BQMs, including the Ising model, have a natural graphical representation. Consider the case for a graph of size $N=4$ with non-zero couplings between the pairs $(s_{0},s_{1})$, $(s_{0},s_{3})$,$(s_{1},s_{2})$, and $(s_{1},s_{3})$. Abstracting from biases and coupling magnitudes, we can represent this model with a graph, $G$, specified by vertices and edges, ($V,E)$. Figure~\ref{fig:graphical_representation} visualizes the graph, where $V = \{s_{0}, s_{1}, s_{2}, s_{3} \}$ and $E = \{ (s_{0},s_{1})$,$(s_{0},s_{3})$,$(s_{1},s_{2})$,$(s_{1},s_{3})\}$.

As we discuss in Section \ref{sec:Methods}, the BQM implied by our problem must be mapped to a graph that is embedded in a QA. While the embedding is typically constructed in a classical pre-processing step -- and, thus, is not a part of the solution -- the problem's difficulty and the embedding's quality depend on the graph our problem implies \citep{Luc14}. Section \ref{sec:Methods} explicitly considers whether a BQM's graph is suitable for a given QA topology.

Earlier, we presented the classical Ising model to enhance intuition. We described the Hamiltonian as an operator that expresses the level of energy in a system as a function of its state and the parameters of the system, which can be configured to encode an optimization problem. In addition, there is a version of the Ising model that allows for quantum phenomena and expresses the level of energy in the system.

To see this, we start with the standard initial state of a QA given by the Hamiltonian:
\begin{equation*}
\mathcal{H}_{0} = -\sum_{i=0}^{N-1} \sigma_{i}^{x},
\label{eqn:quantum_annealer_initial}
\end{equation*}
where $\sigma_{i}^{x} = I \otimes ... \otimes \sigma^{x} \otimes ... \otimes I$, $I$ is a 2x2 identity matrix, and $\sigma^{x}$ is the Pauli $X$ matrix:
\begin{equation*}
\sigma^{x} = 
\begin{bmatrix}
0 & 1 \\
1 & 0
\end{bmatrix},
\label{eqn:pauli_x}
\end{equation*}
where the $i$ indicates that $\sigma^{X}$ is located in the $i^{th}$ position in the sequence of tensor products:

\noindent For the $N=2$ case, for example, $\mathcal{H}_{0}$ would be specified as:
\begin{equation*}
    \mathcal{H}_{0} = - (\sigma_{x} \otimes I + I \otimes \sigma_{x}) = \begin{bmatrix}
0 & -1 & -1 & 0 \\
-1 & 0 & 0 & -1 \\
-1 & 0 & 0 & -1 \\
0 & -1 & -1 & 0 \\
\end{bmatrix}.
\label{eqn:quantum_annealer_initial_N2}
\end{equation*}

\noindent The ground state corresponds to the eigenvector associated with the minimum eigenvalue of $\mathcal{H}_{0}$. In this case, the characteristic polynomial of the expression for $\mathcal{H}_{0}$ is $\lambda^{4} - 4 \lambda = 0$, yielding the eigenvalues $\lambda = \{-2, 2, 0, 0\}$ and eigenvectors $v$:
\begin{equation*}
v = \left\{
\begin{bmatrix}
1 \\
1 \\
1 \\
1  \\
\end{bmatrix},
\begin{bmatrix}
1 \\
-1 \\
-1 \\
1  \\
\end{bmatrix},
\begin{bmatrix}
-1 \\
0 \\
0 \\
1 \\
\end{bmatrix},
\begin{bmatrix}
0 \\
-1 \\
1 \\
1  \\
\end{bmatrix}
\right\}.
\label{eqn:eigenvectors_N2}
\end{equation*}

\noindent The first eigenvector, $v_{0} = [1,1,1,1]$, is associated with the minimum eigenvalue ($\lambda_{0} = -2$) and, thus, corresponds to the global solution. With a suitable normalization, this vector represents a uniform superposition over all possible states. We have two qubits and two states in this case, so there are four possible classical configurations. A QA is typically prepared in this position (but for $N$ qubits), since it has a trivially computable ground state. Performing measurement immediately would collapse the superposition, yielding one of the four possible states with equal probability.

During the annealing process, the Hamiltonian can be expressed as:
\begin{equation}
    \mathcal{H} = \left(1-\frac{t}{T} \right)\sum_{i=0}^{N-1} h_{i} \sigma^{x}_{i} + \frac{t}{T}\sum_{i=0}^{N-1} \sum_{j=0}^{N-1} J_{i,j} \sigma^{z}_{i} \sigma^{z}_{j},
\label{eqn:quantum_ising_anneal}
\end{equation}
where the coefficients on the initial and problem Hamiltonians correspond to the annealing schedule. Again, we use $\sigma_{z}^{i}$ as a shorthand for $\sigma_{i}^{z} = I \otimes ... \otimes \sigma^{z} \otimes ... \otimes I$, where $\sigma_{z}$ is the Pauli $Z$ matrix:
\begin{equation*}
\sigma^{z} = 
\begin{bmatrix}
1 & 0 \\
0 & -1
\end{bmatrix}.
\label{eqn:pauli_z}
\end{equation*}

The Hamiltonian \eqref{eqn:quantum_ising_anneal} contains \textit{transverse fields}, represented by the $\sigma^{x}_{i}$ terms, which pull qubits toward a superposition of the +1 and -1 states. These fields are in tension with the $\sigma_{z}^{i}$ terms in the problem Hamiltonian, which pull qubits toward a classical computational basis state.\footnote{The eigenvectors of $\sigma^{z}$ are $[1,0]$ and $[0,1]$, which correspond to +1 and -1 in the classical Ising model or 0 and 1 in the computational basis. The eigenvectors of $\sigma^{x}$ are $[1/\sqrt{2}, 1/\sqrt{2}]$ and $[1/\sqrt{2}, -1/\sqrt{2}]$, which are equal superpositions of the classical Ising states.}

\paragraph{The QUBO model.} The QUBO model is still a BQM, but it uses 0 and 1 states rather than the +1 and -1 states of the Ising model. A QUBO model allows equality and inequality constraints and does not use a quadratic approximation.\footnote{The QUBO model allows for the use of constraints, but they must be incorporated into the objective function. Additionally, the term \textit{quadratic} refers to the number of interactions between binary variables, not to the more familiar concept of a quadratic approximation in computational economics. As we will see, this is also not limited to including higher-order terms, which can be added into a QUBO after \textit{quadratization}.}  

The standard form of a QUBO problem is:
\begin{equation*}
    \mathcal{H}(x) = \sum_{i}^{N} Q_{i,i} x_{i} + \sum_{j < i}^{N} Q_{i, j} x_{i} x_{j}.
\label{eqn:qubo}
\end{equation*}
This Hamiltonian formulation emphasizes the relationship between the Ising model and QUBO problems in the context of quantum annealing. 

Notice that $x = \{ x_{0},x_{1},...,x_{N-1} \}$ is a vector of binary variables that can take on values of 0 and 1. The problem structure is embedded in $Q$, an upper-triangular matrix. We may rewrite the problem in terms of $Q$ as $\mathcal{H}(x) = x^{T} Q x$.

\paragraph{Model conversion.} Converting the constrained optimization problem implied by an economic model into a BQM is not trivial. Hence, as discussed next, we will take the intermediate step of mapping the objective function to a pseudo-Boolean function (PBF). We will then explain how to decompose the PBF into a QUBO and expand the objective function to include constraints.

\subsubsection{Pseudo-Boolean functions}\label{sssec:pbfs}

NP-hard problems must typically be converted to a PBF before they can be reduced to a QUBO \citep{VCML18}. This includes most optimization problems in economics and finance that could be solved on a quantum annealer. A PBF takes the form $f:\mathcal{B}^{N} \rightarrow \mathbb{R}$, where $\mathcal{B}$ is the Boolean domain.

\citet{BH02} show that each PBF has a unique representation as a multi-linear polynomial:
\begin{equation*}
    f(x) = \sum_{\mathcal{M} \subseteq \{ 1,...N \}} \alpha_{\mathcal{M}} \prod_{i \in \mathcal{M}} x_{i},
\label{eqn:multi_linear_poly}
\end{equation*}
which can be reduced to a QUBO in polynomial time and with a polynomial bound on its size.

Thus, if we can map an optimization problem to a PBF, we can also reduce it to a QUBO and potentially solve it on a QA. This is not obvious since most challenging problems in economics do not have a clear mapping to a QUBO but may be reformulated as a PBF.

\paragraph{Partial example of PBF mapping.} For illustration, consider an optimization problem that entails selecting next-period capital, $k'$, given the current capital stock, $k$, and a productivity shock, $z$. If we want to solve a discrete approximation to the problem, one approach is to define a grid for $z$ and a shared grid for $k$ and $k'$, where each maps an index to a value for a given variable. We can then construct a loss function that takes the model variables' indices, $(i,j,m)$, and returns a scalar-valued loss.

If we wanted to reconstruct the loss function as a QUBO, we would encounter two problems. First, the loss function requires us to specify the indices of three variables, but QUBOs are restricted to interactions between two binary variables. And second, the loss function takes integer-valued indices as inputs, but a QUBO specifies states in terms of binary variables.

An inefficient candidate solution to the second problem is to map each node in the grids for $k$, $z$, and $k'$ to binary variables. That is, if $k$ is at node $j$ in its grid, then $x_{k_{j}} = 1$ and $x_{k_{\not j}} = 0$. Thus, we can represent the variable indices $(i,j,m)$ from the original problem as $x_{k_{i}}x_{z_{j}}x_{k'_{m}} = 1$ in a PBF. However, we cannot include such a term in a QUBO since it consists of the product of three variables. One solution to this problem is to quadratize the term: that is, rewrite it in terms of quadratic and lower-order terms. Next, we show how to do this.

\subsubsection{Degree reduction}\label{ssec:degree_reduction}

Once we have mapped the initial problem to a PBF, the next step is to perform degree reduction until all non-quadratic terms are eliminated. A good quadratization strategy should yield a QUBO with the same global minimum as the PBF and minimize the number of auxiliary variables introduced.

\citet{Dat19} and \citet{DC19} evaluate more than 30 methods of quadratization. In Online Appendix \ref{sec:quadratization_methods}, we discuss four methods that introduce either no auxiliary variables or the minimum number of auxiliary variables. In the text below, we introduce a method that can be applied in all circumstances but with less appealing properties. All methods are selected for their suitability for use on D-Wave's \textit{Pegasus} QA topology (the one employed by our application).

\paragraph{Reduction by substitution.} A general strategy for performing degree reduction is to replace products of two variables in higher-order terms with an auxiliary variable and a quadratic constraint. Consider, for instance, the PBF:
\begin{equation*}
    \mathcal{H}(x) = 2x_{1}x_{2}x_{3} + 4x_{2}x_{3}x_{4} - 5x_{2}x_{3}x_{5}.
\end{equation*}

\noindent Since the product $x_{2}x_{3}$ is common to all monomials, performing the substitution $x_{a}=x_{2}x_{3}$ reduces all cubic terms to a quadratic form. However, it also requires the introduction of a quadratic constraint, yielding the modified form of the Hamiltonian:
\begin{equation*}
    \mathcal{H'}(x) = 2x_{1}x_{a} + 4x_{a}x_{4} - 5x_{a}x_{5} + \gamma \underbrace{(x_{2}x_{3} - 2(x_{2}+x_{3})x_{a} + 3x_{a})}_{x_{a}=x_{2}x_{3}}.
\label{eqn:substitution_2}
\end{equation*}
The penalty, $\gamma >0$, is a hyperparameter that must be determined by the researcher.

While this approach provides a general-purpose strategy for degree reduction, it comes at a high cost: one auxiliary variable and one penalty parameter for each degree reduced. The penalty parameter is especially problematic since QAs are sensitive to the scale of the largest model parameter. Thus, introducing a sufficiently large $\gamma$ to impose the constraint will weaken the relative strengths of couplings related to the problem itself.

\subsubsection{Adding constraints}\label{ssec:adding_constraints}

The \textit{unconstrained} part of the term quadratic unconstrained binary optimization refers to constraints \textit{outside} of the QUBO problem's objective, which are not permitted. It is, however, possible to introduce constraints into the QUBO problem. For example, consider the case where we have the objective function $f(x) = x_{1}x_{2}x_{3} + 3x_{1}x_{3} + 2x_{2}$ and the constraint $x_{1}+x_{2} = 1$. We can express this problem as an objective function with a penalty term, as in $\mathcal{H}(x) = x_{1}x_{2}x_{3} + 3x_{1}x_{3} + 2x_{2} + \gamma (1-x_{1}-x_{2})^{2}$.

\subsection{Computational Complexity}\label{ssec:computational_complexity}

Problems solved on a QA are mapped to a physical quantum Ising problem, which is NP-hard to solve exactly on a classical computer \citep{Bar82}. Additionally, the associated \textit{decision problem} is NP-complete, which implies that the problem can be mapped to any other NP problem in a polynomial number of steps.

Consequently, if a QA could solve the Ising problem exactly in polynomial time, it would also be able to solve any NP problem in polynomial time. However, the theoretical and experimental evidence suggests that this is unlikely to be the case. Rather, it appears that QAs can find high-quality approximation solutions fast but can only slowly provide further improvements in the quality of the solution \citep{ZBT+15, KYN+15}.

QAs are unlikely to solve the Ising problem in polynomial time because the transition between the initial and problem Hamiltonian must be adiabatic to ensure that the system remains in the ground state. Faster transitions can lead to a jump in the energy level from the ground state to the first excited state, which can prevent the system from terminating in a state consistent with the global minimum at the end of the annealing process.

Thus, while the inability to perform the transition adiabatically prevents us from guaranteeing that we remain in the ground state, it does not rule out the possibility that we end in either the ground state or another low-energy state. In cases where the spectral gap remains large throughout the transition process, a QA may transition sufficiently slowly to ensure that this happens.

\begin{figure}[htb!]
\begin{center}
\resizebox{0.5\textwidth}{!}{
\begin{tikzpicture}
\begin{axis}[
    ticks = none,
    axis lines = left,
    legend pos=north west,
    xlabel = {Input Size},
    ylabel = {Complexity},
]
\addplot [
    domain=0:16, 
    samples=20, 
    color=black,
    mark=o,
]
{2^x};
\addlegendentry{$\mathcal{O}(2^{x})$}
\addplot [
    domain=0:16, 
    samples=20, 
    color=black,
    mark=square,
    ]
{100*x^2};
\addlegendentry{$\mathcal{O}(x^{2})$}
\addplot [
    domain=0:16, 
    samples=20, 
    color=black,
    style=dashed,
    ]
{100*10^2};
\end{axis}
\end{tikzpicture}
}
\end{center}
\caption{Comparison of the computational complexity of algorithms with polynomial and exponential time complexity for different input sizes.}
\label{fig:problem_complexity}
\end{figure}
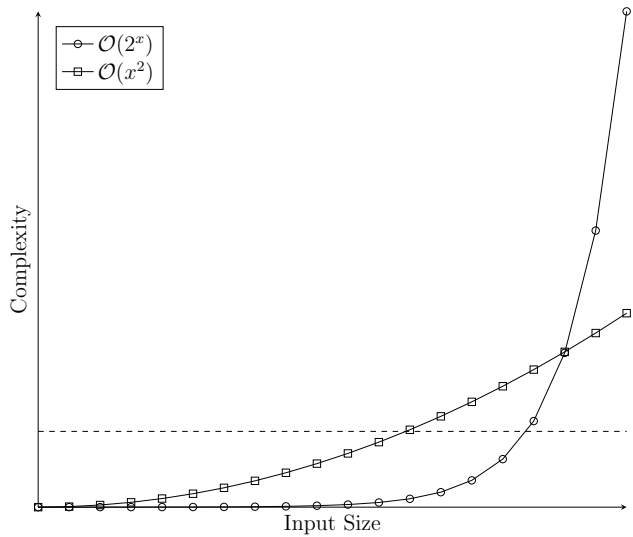

More precisely, the inability to guarantee that an annealing problem can be reduced to polynomial complexity might not preclude a substantial quantum advantage. Consider the case where a problem with exponential time complexity is reducible to one with polynomial time complexity, but in both cases, the computational resource requirements are sufficiently high that only small problem instances can be solvable. As depicted in Figure \ref{fig:problem_complexity}, the exponential time solution could solve larger problems. The line with circular markers represents an algorithm with exponential complexity, and the line with square markers represents an algorithm with polynomial complexity. The dashed line indices the maximum permissible complexity given available computational resources. Thus, even if a QA requires exponential time, it can considerably speed up feasible problem cases.

\subsection{Limitations}\label{ssec:limitations}

Quantum computers enable us to execute algorithms that require less time and space than the best available classical algorithms for certain problem classes. But they also place unfamiliar limitations on the computations that can be performed. We discuss the most important limitations in this subsection.

\paragraph{Quantum-to-classical bottleneck.} QAs use quantum superposition, which enables qubits to be in a linear combination of 0 and 1 states rather than in the 0 or 1 state, as with classical bits. Similarly, two bits on a classical computer can be in either the 00, 01, 10, or 11 configurations. In contrast, a pair of qubits on a QA can be in a linear combination of all four states simultaneously. More generally, for the $N$-qubit case, quantum superposition enables the system to be in a linear combination of $2^{N}$ states.

Superposition allows us to start a QUBO problem in a uniform linear combination over all $2^{N}$ states rather than in an individual state.\footnote{If there are $2^{N}$ states, then the degree to which we will be in each of those states is $\frac{1}{2^{N}}$. If, for instance, we perform measurement on the system to see which state we are in, the superposition will collapse, and a random state will be chosen from a uniform distribution over the $\frac{1}{2^{N}}$ states.} During the annealing process, the system transitions from a quantum superposition state to a pure or basis state, which solves our problem. The QA then performs measurement and reads out classical bits corresponding to that state.

The fact that we cannot observe a quantum state directly and must instead sample from it using measurement is referred to as the \textit{quantum-to-classical bottleneck}. Even if we create a quantum superposition over $2^{N}$ states, we can only retrieve $N$ classical bits of information from each annealing step. Current state-of-the-art QAs have 5000 qubits, which means that each annealing step can return at most 0.625kB of classical information. Consequently, quantum annealing applications typically center around problems with a parsimonious solution.

\paragraph{Classical-to-quantum overhead.} The time to perform the quantum annealing step is at least $5\mu s$ (microseconds) but is independent of the problem size. In addition to this, programming the quantum processing unit (QPU) takes an additional $9ms$ (milliseconds), and reading out the classical output of a project takes an additional $120 \mu s$ \citep{VCML18}. Thus, the total run time is $T = T_{p} + R(T_{a} + T_{r})$, where $T_{p}$ is the programming time, $T_{a}$ is the anneal time, $T_{r}$ is the readout time, and $R$ is the number of repetitions.

We perform the anneal and readout $R$ times but only program the QPU once. Furthermore, the programming step takes 72 times as long as the anneal and readout are combined. As a result, problems that need partitioning and solving using a hybrid algorithm that employs both a classical CPU and a QPU suffer from a substantial classical-to-quantum transfer overhead.

Since quantum annealing is a probabilistic process, it may yield a different result for each repetition. For this reason, we will typically perform multiple anneals and readouts for each problem ($R>1$), allowing us to identify the frequency and energy level of different solutions. This will have the effect of reducing the importance of the transfer overhead.

\paragraph{Noise.} Unlike classical computers, quantum computers cannot yet perform error correction or store information in memory for extended periods because of 1) the impossibility of copying quantum states \citep{WZ82}; and 2) the natural tendency of quantum systems to decohere in nanoseconds or microseconds. In this respect, QAs have an advantage over universal quantum computers since the annealing algorithm is less affected by qubit decoherence \citep{CFP+01,AL15}.

\begin{figure}[htbp!]
\minipage{0.42\textwidth}
  \includegraphics[width=\linewidth]{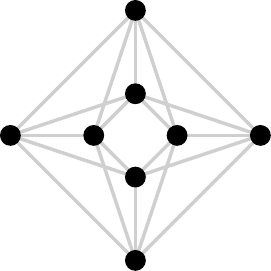}
  \caption*{Intra-Cell Connections}
\endminipage\hfill
\minipage{0.42\textwidth}
  \includegraphics[width=\linewidth]{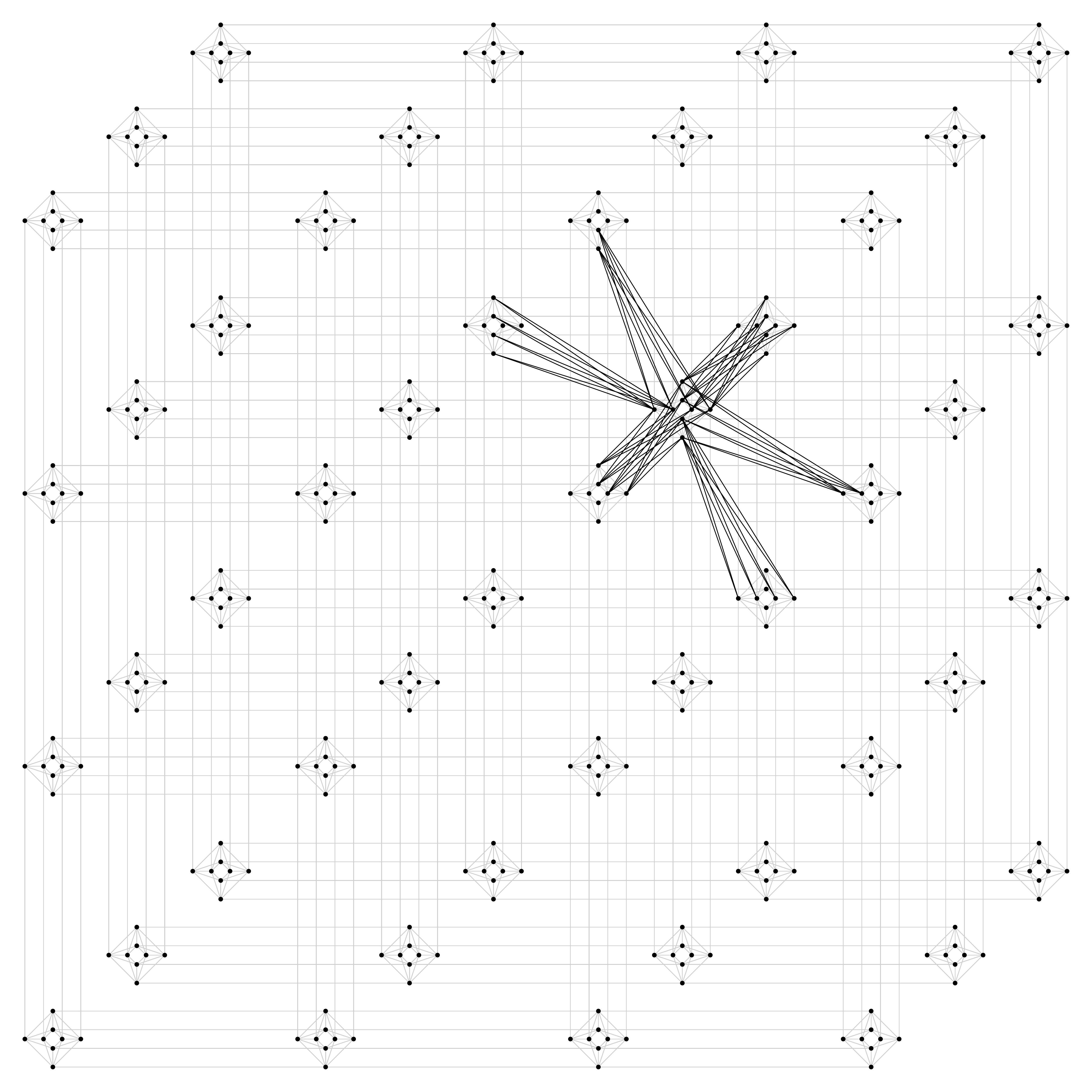}
  \caption*{Inter-Cell Connections}
\endminipage
\caption{Intra-cell and inter-cell connections on subgraphs of the 16x16x3 \textit{Pegasus} topology. The left panel shows the connections within an eight-qubit cell. The right panel shows the inter-cell connections for an example cell embedded in a 4x4x3 subgraph. The figure was generated using the approach in \citet{DSC19}.}
\label{fig:pegasus_topology}
\end{figure}

\paragraph{QPU topology.} Once we have reduced our initial problem to a QUBO, we must embed it in the QPU's graph. So, the graph topology will place another set of restrictions on the types of problems we can solve with a QA. Figure \ref{fig:pegasus_topology} shows a cell and a subgraph from the \textit{Pegasus} topology, which is used in D-Wave's \textit{Advantage} QAs.

The primary topological considerations include the total number of qubits, the total number of couplers, and the number of couplers per qubit. Having fewer qubits means we can represent fewer binary variables in our QUBO and read out less classical information. Similarly, a lower total number of couplers means we will be less able to use \textit{quantum entanglement} as a computational resource, which is needed to specify the strength and direction of relationships between variables.

In addition to the total number of couplers, the number of couplers per qubit will also be important, as it will determine the extent to which \textit{chaining} is needed. Chaining occurs when qubits are not directly coupled but have a quadratic term in the QUBO. It entails creating an indirect link by forcing the qubits between them to have the same state as one of the qubits in the quadratic term. Long chain lengths reduce the number of qubits available to represent the problem and force us to tune a \textit{chain strength} hyperparameter.

\subsection{Ideal Problems}\label{ssec:ideal_problems}

Given the features of a QA, what constitutes an ideal problem that could demonstrate quantum advantage? The literature suggests that the following three problem attributes are important: 1) a large state space and parsimonious solution; 2) long, thin barriers between minima in the energy landscape (loss function); and 3) opportunities for reducing the loss through co-tunneling.  

\paragraph{Large state-space and parsimonious solution.} A problem that is solved directly on a QPU and cannot be partitioned into subproblems should have a solution that requires no more than $N$ bits to express, where $N$ is the number of qubits. D-Wave's \textit{Advantage} line of QAs, for example, have more than 5000 qubits and can output 0.625kB of classical information per anneal. An ideal problem would exploit quantum superposition by traversing a large state-space but should ultimately have a solution expressible using a small amount of classical information. A challenging equilibrium-finding problem, for instance, would be a good candidate.

\begin{figure}[htpb!]
\centering
\includegraphics[width=0.50\linewidth]{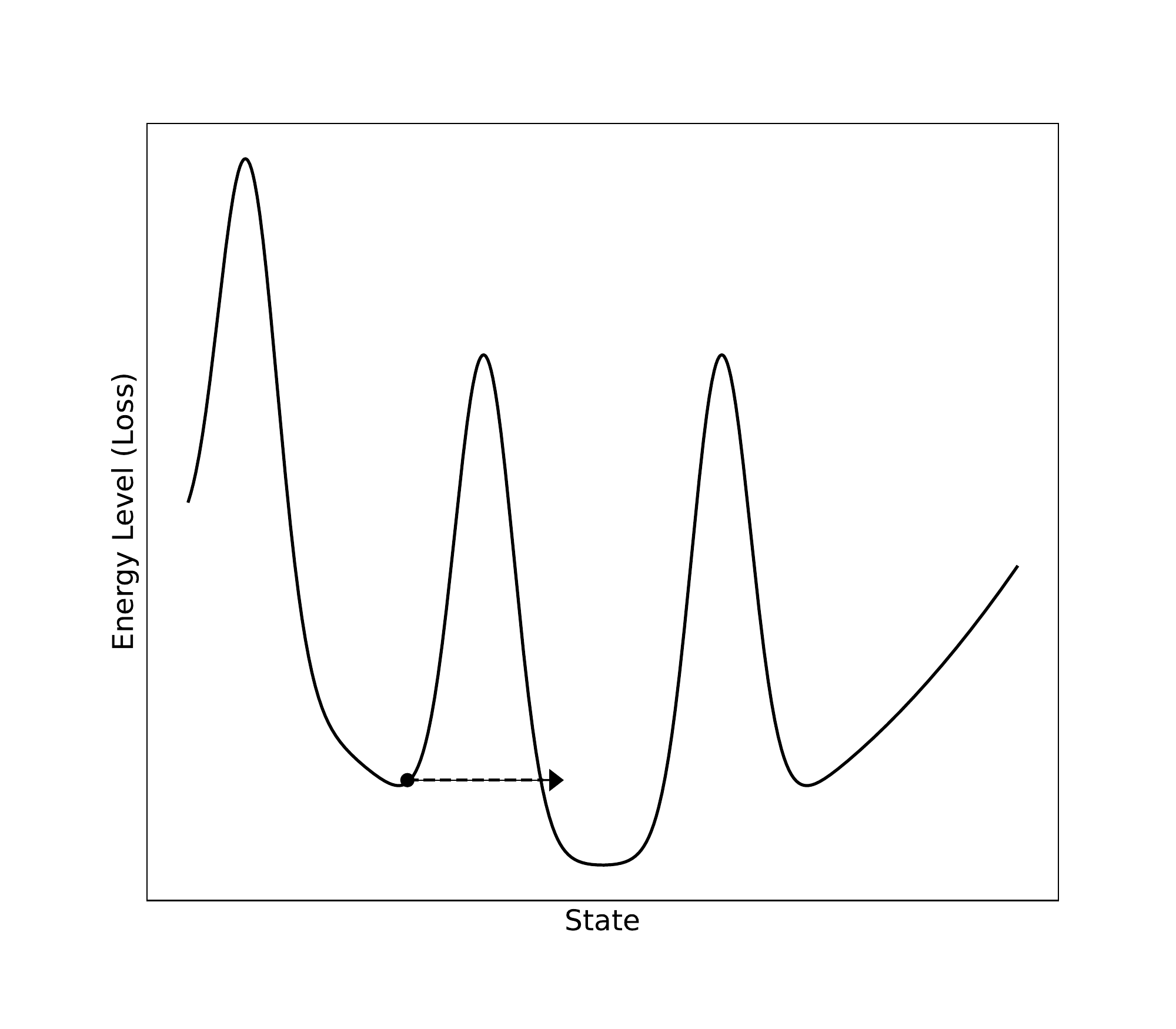}
\caption{\textit{Quantum tunneling} is depicted by the dashed line passing through the energy landscape.}
\label{fig:quantum_tunneling}
\end{figure}

\paragraph{Barriers between minima in the loss function.} QAs exhibit \textit{quantum tunneling}, where a qubit passes through a barrier in the energy landscape \citep{BRI+14,AVM+15,ARTL15}, rather than attempting to climb over it, as a classical optimization algorithm would do. \citet{DBI+16} suggest that problems with loss functions that have long, thin barriers surrounding local minima are ideal candidates for tunneling. Figure \ref{fig:quantum_tunneling} depicts the loss function for a problem that may permit a tunneling-induced speed-up. Classical solvers will get stuck in local minima, whereas QAs may tunnel through the barriers to the global minimum.

\paragraph{Loss reduction through co-tunneling.} \citet{DBI+16} also exploit \textit{co-tunneling} \citep{BSS+16}, where multiple entangled qubits jointly tunnel through a barrier to a lower energy level (loss). If the qubits were not entangled, it would not be possible for all of them to tunnel through the barrier instantaneously. Ideal uses of co-tunneling involve areas in the loss function where changing one variable at a time in a particular direction will increase the loss, but changing all simultaneously will decrease it.

\section{The Model}\label{sec:Model}

To show the potential of QA in economics, we solve the version of the real business cycle (RBC) model used in \AFV, a well-known testbed for computation economics. A social planner chooses a sequence of consumption $c_{t}$ and capital $k_{t+1}$ to solve:
\begin{equation*}
    \underset{\{ c_{t},k_{t+1} \}}{\text{max}} \mathbb{E}_{0} \sum_{t=0}^{\infty} (1-\beta)\beta^{t} \ln(c_{t})
\label{eqn:objective_fcn}
\end{equation*}
subject to the resource constraint $c_{t} + k_{t+1} = z_{t} k_{t}^{\alpha} + (1-\delta)k_{t}$, where $\beta$ is the discount factor, $z_t$ is a productivity shock, and $\mathbb{E}_{0}$ is the conditional expectations operator. 

To match the comparison exercise in \AFV{} and get a closed-form solution to benchmark our QA solution, we set $\beta=0.95$, $\alpha=0.33$, and $\delta=1$. Recall that, with full depreciation, the optimal decision rules for the social planner are given as $c_{t} = (1-\alpha \beta)z_{t}k_{t}^{\alpha}$ and $k_{t+1} = \alpha \beta z_{t} k_{t}^{\alpha}$. Also as in \AFV{}, $z_{t}$ is a Markov chain with support $z_{t} \in \{0.9792, 0.9896, 1.0000, 1.0106, 1.0212\}$ and transition matrix:
\begin{equation*}
\Pi =
\begin{pmatrix}
0.9727 & 0.0273 & 0 & 0 & 0 \\
0.0041 & 0.9806 & 0.0153 & 0 & 0 \\
0 & 0.0082 & 0.9837 & 0.0082 & 0 \\
0 & 0 & 0.0153 & 0.9806 & 0.0041 \\
0 & 0 & 0 & 0.0273 & 0.9727 \\
\end{pmatrix},
\label{eqn:transition_matrix}
\end{equation*}
which approximates an AR(1) process for log productivity.

\AFV{} write the model in the previous section as a Bellman equation:
\begin{equation*}
    v(k,z) = \underset{k'}{\text{max }} \{ \ln(zk^{\alpha} - k') + \beta \mathbb{E} [v(k',z')|z] \}
\label{eqn:value_function}
\end{equation*}
and discretize the capital stock into 17,820 uniformly spaced points over $[0.5\bar{k},1.5\bar{k}]$, where $\bar{k}$ is capital at the steady state. Then, they solve the problem using value function iteration (VFI). In \texttt{C++}, \AFV{} report a computation time of 0.73 seconds (on a 2023 vintage PC, the running time would be around 0.4 seconds).

We cannot implement VFI on current QAs. State-of-the-art machines currently output 0.625kB of classical information per anneal, while the look-up table from the above dynamic programming problem requires 89,100 floating-point numbers, at least 285 times as much classical information, even if we use half-precision.

We get around this problem by using parametric dynamic programming (PDP), which allows for a representation of the solution that can be encoded in a small number of bits.\footnote{We will evaluate the value function at all productivity nodes and at a number of capital nodes proportional to the square root of the number of nodes in \AFV{}. This is already more than the number needed to identify the value function parameters. The parameters used to represent the value function in the quantum algorithms are selected from a set that has between $2^{14}$ and $2^{20}$ elements, depending on the algorithm.} Furthermore, even if there were no quantum-to-classical bottleneck, PDP would still be preferable, since our intention is to eventually solve large problem instances on a quantum annealer. Even if we could output an exponential amount of classical information for a large problem instance, it would take an exponential amount of time and could not be stored on a classical computer. 

In applications, PDP tends to yield solutions faster than VFI and is easier to scale up in terms of dimensionality of the state space, but it is often less stable (see, e.g., \citealp{TU90}, \citealp{BRG+00}, and \citealp{Swe13}) and does not have a provably lower time complexity. Nonetheless, to ensure full compatibility of results, we will solve our PDP problem first on a classical computer to establish a clear benchmark.

\section{Methods}\label{sec:Methods}

We now describe how to set up the PDP problem associated with our RBC model, translate it into a PBF, quadratize the PBF into a QUBO, and execute the QUBO on a QA.

\subsection{Setting up the PDP}\label{ssec:methods_PDP}

We use a parametric policy iteration (PPI) formulation that follows \citet{BRG+00} where both the policy and value functions have parametric forms to ensure they are expressed parsimoniously. PPI also typically yields solutions faster than discrete policy iteration but may have worse convergence properties.

We start by assuming functional forms for the value function and the consumption and capital policy rules:
\begin{gather}
    v(k,z) = x_{2} + x_{3}\ln(y).
    \label{eqn:value_function_pdp}\\
    k' = x_{1} y
    \label{eqn:policy_fcn_capital}\\
    c = (1-x_{1}) y.
    \label{eqn:policy_fcn_cons}
\end{gather}
Then, we take the Bellman equation $v(k,z) = \underset{k'}{\text{max}} \{ \ln(c) + \beta \mathbb{E}[v(k',z')|z] \}$, and use equations \eqref{eqn:policy_fcn_capital} and \eqref{eqn:policy_fcn_cons} to rewrite it as:
\begin{equation}
    v(k,z) = \underset{x_{1}}{\text{max}} \{ \ln[(1-x_{1})y] + \beta \mathbb{E}[v(x_{1}y,z') | z] \}.
    \label{eqn:value_function_pdp_1}
\end{equation}

In more complex models, one would need to specify richer functional forms. For instance, neural networks could be used to approximate arbitrary value and policy functions \citep{fernandez2020}. While this would require more parameters $x_1$, $x_2$, ..., (i.e., the weights of the neural network), it is conceptually straightforward.

PPI alternates between policy improvement and policy valuation steps. The policy improvement step uses equation \eqref{eqn:value_function_pdp} and is given by:
\begin{equation*}
    x_{1}^{*} = \underset{x_{1}}{\text{argmin}} \{ -\ln[(1-x_{1})y] - \beta \mathbb{E}[\bar{x}_2 + \bar{x}_3\ln(y') | z] \}.
    \label{eqn:policy_update_step}
\end{equation*}
Notice that $\bar{x}_{2}$ and $\bar{x}_3$ are taken as fixed parameters in this step, as indicated by the bar.

Re-arranging equation \eqref{eqn:value_function_pdp_1} yields a residual function. The policy valuation step minimizes the sum of the squared residuals by choosing $x_{2}$ and $x_{3}$:
\begin{equation*}
    x_{2}^{*}, x_{3}^{*} = \underset{x_{2},x_{3}}{\text{argmin}} \{ \ln[(1-\bar{x}_{1})y] + \beta \mathbb{E}[x_{2} + x_{3}\ln(y') | z] - x_{2} - x_{3} \ln(y) \}^{2}.
    \label{eqn:value_update_step}
\end{equation*}

To simplify notation, we denote the objective function for the policy improvement step as $g_{p}(x_{1};\bar{x}_{2},\bar{x}_{3})$ and the objective function for the policy valuation step as $g_{v}(x_{2},x_{3};\bar{x}_{1})$. Given some initial parameters $\{ x_{1}, x_{2}, x_{3} \}$, PPI iterates on $g_{p}(x_{1};\bar{x}_{2},\bar{x}_{3})$ and $g_{v}(x_{2},x_{3};\bar{x}_{1})$ until achieving convergence. Algorithm \ref{alg:classical_ppi} summarizes the previous steps.

\begin{algorithm}
\DontPrintSemicolon
Select a parametric form for the policy and value functions.\;
Initialize the parameters of the policy function, $x_{1}$, and value function, $\{ x_{2}, x_{3} \}$.\;
Update the policy function parameter: $x_{1}^{*} = \underset{x_{1}}{\text{argmin}} \{ g_{p}(x_{1};\bar{x}_{2},\bar{x}_{3}) \}$.\;
Update the value function parameters: $x_{2}^{*}, x_{3}^{*} = \underset{x_{2},x_{3}}{\text{argmin}} \{ g_{v}(x_{2},x_{3};\bar{x}_{1}) \}$ \;
Repeat steps 3 and 4 until the convergence criterion is satisfied.
\caption{Parametric Policy Iteration}\label{alg:classical_ppi}
\end{algorithm}

\begin{figure}[!htb]
\includegraphics[width=\linewidth]{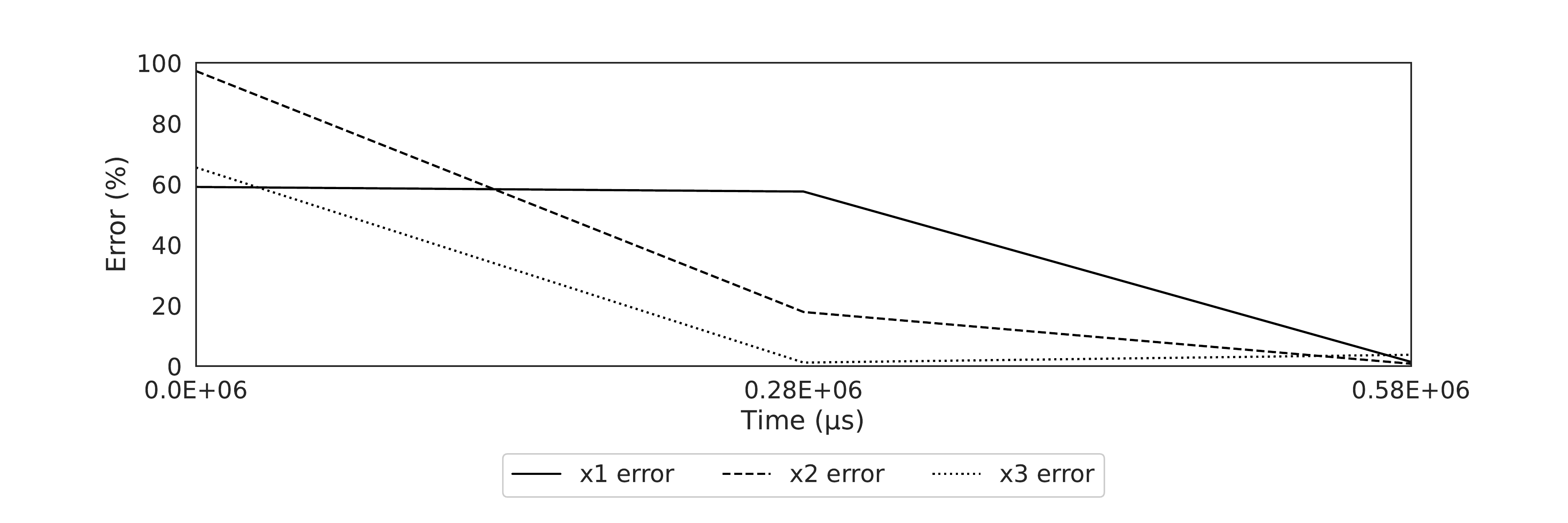}
\caption{Policy and value function parameter errors as a function of average execution time over 10 runs. Convergence is achieved after two iterations.}
\label{fig:classical_solution}
\end{figure}

Coding Algorithm \ref{alg:classical_ppi} on a classical computer is straightforward. Figure \ref{fig:classical_solution} plots the average results from 10 executions of Algorithm \ref{alg:classical_ppi} on a classical computer. The horizontal axis shows the total execution time in microseconds. Each tick corresponds to an iteration. The vertical axis shows the parameter errors in absolute percentage deviations from their true values (which, in our version of the RBC model, we can compute analytically). The terminal errors for $x_{1}$, $x_{2}$, and $x_{3}$ are 1.37\%, 0.77\%, and 3.74\% respectively. Convergence is achieved after 0.58 seconds, which is faster than the 0.73 seconds required for the \texttt{C++} benchmark VFI solution, but slightly slower than the 0.40 seconds required for the same benchmark solution on 2023-era hardware.

Therefore, our execution time comparisons with the \AFV{} benchmarks later in the paper are not biased in favor of the QA because of the use of PPI instead of VFI; however, we will show how a QA can deliver solutions one order of magnitude faster than a classical computer, regardless of whether the latter uses PPI or VFI.

Programming a quantum annealer to execute Algorithm \ref{alg:classical_ppi} is more challenging since we must first re-express it as a QUBO problem. This transformation requires the intermediate step of formulating it as a pseudo-Boolean optimization (PBO) problem. The next two subsections show how we can implement these steps.

\subsection{Translating the PDP into a PBO}
\label{ssec:methods_PBF}

Formulating the PDP problem as a pseudo-Boolean optimization (PBO) problem entails first mapping the parameters $\{ x_{1}, x_{2}, x_{3} \}$ to binary variables. The simplest reformulation would discretize the parameter values, representing each node as a binary variable, and impose the condition that the sum of the binary variables equals one. However, this would be inefficient and soon exhaust the QA's computational resources. 

An alternative strategy, which we will use to represent $\{ x_{2}, x_{3} \}$, re-expresses each variable using a standard binary encoding. This strategy allows us to represent $N$ nodes $x_{\nu}$ in the state space using $\lceil \log_{2}(N) \rceil$ qubits:
\begin{equation*}
    x_{\nu} = s_{\nu} \sum_{j_{\nu}=0}^{J_{\nu}} 2^{j_{\nu}} x_{\nu, j_{\nu}}.
    \label{eqn:x_binary}
\end{equation*}
Notice that $s_{\nu}$, where $\nu \in \{1, 2, 3\}$, is the scaling factor for variable $x_{\nu}$, defined such that $x_{\nu} \in [0,s_{\nu}(2^{J_{\nu}+1}-1)]$. 

As a minimal example, consider the case for $\nu = 2$, where $s_{2}=1$ and $J_{2}=2$. If $x_{2, 0} = 1$, $x_{2, 1} = 0$, and $x_{2,2}=1$, then $x_{2} = 2^{0}*1 + 2^{1}*0 + 2^{2}*1 = 5$. If, instead, we want to set the maximum value to $m$, then we can impose the following scaling factor: $s_{2} = \frac{m}{2^{J_{2}+1}-1}$. Raising $J_{2}$ increases the precision of $x_{2}$ for a fixed $s_{2}$. Using just 19 qubits, we can construct a grid over $x_{2}$ values with more than $10^6$ nodes that does not require any quadratic terms or constraints.\footnote{The difference in size between the coefficient on the smallest and largest terms will become problematic when using a binary variable encoding on a QA. Thus, it is still worthwhile to consider how much precision is needed for a given problem.}

A second, more substantial, challenge is representing the terms $\ln(x_{1})$ and $\ln(1-x_{1})$ from Algorithm \ref{alg:classical_ppi} on a QA. A naive approach might treat $\ln(x_{1})$ and $\ln(1-x_{1})$ as separate variables and link them with a constraint. However, this approach is inefficient, as it requires the use of two sets of binary variables to represent $\ln(x_{1})$ and $\ln(1-x_{1})$. It also requires us to invert $\ln(x_{1})$ and $\ln(1-x_{1})$ and impose a quadratic constraint, which substantially increases the connectivity requirements and network size.

Instead, we use a single set of binary variables to represent $x_{1}$ and two sets of coefficients to transform it into $\ln(x_{1})$ and $\ln(1-x_{1})$ using polynomial approximations:
\begin{gather}
    \ln(x_{1}) \approx \Bigg[a_{0} + \sum_{j_{1}=0}^{J_{1}} a_{1} 2^{j_{1}} x_{1, j_{1}}
    + a_{2} \sum_{j_{1}=0}^{J_{1}} \sum_{i_{1} = 0}^{J_{1}} 2^{j_{1} + i_{1}} x_{1, j_{1}} x_{1, i_{1}} \Bigg]
    \label{eqn:x1_approx_1}\\
    \ln(1-x_{1}) \approx \Bigg[\tilde{a}_{0} + \tilde{a}_{1} \sum_{j_{1}=0}^{J_{1}} 2^{j_{1}} x_{1, j_{1}} \Bigg],
    \nonumber
\end{gather}
where $\{ a_{0},a_{1},a_{2} \}$ and $\{ \tilde{a}_{0},\tilde{a}_{1} \}$ are the sets of coefficients.\footnote{We have $[a_{0}, a_{1}, a_{2}] =  [-0.10905, 0.57570, -1.38445]$ and $[ \tilde{a}_{0},\tilde{a}_{1} ] = [-0.22278, -0.28375]$.}  

Then, we can exploit symmetry and the properties of binary variables (i.e., $x^{2} = x$) to rewrite equation \eqref{eqn:x1_approx_1} as:
\begin{equation*}
    \ln(x_{1}) \approx \Bigg[a_{0} + \sum_{j_{1}=0}^{J_{1}} (a_{1} 2^{j_{1}} + 
    a_{2} 2^{2j_{1}}) x_{1, j_{1}}
    + 2 a_{2} \sum_{j_{1}=0}^{J_{1}} \sum_{i_{1} < j_{1}} 2^{j_{1} + i_{1}} x_{1, j_{1}} x_{1, i_{1}} \Bigg].
    \label{eqn:x1_approx_2}
\end{equation*}
This formulation is useful because it reduces the number of auxiliary variables that must be introduced to quadratize cubic and higher-order terms.

The use of quadratic terms does not require the introduction of additional binary variables. Hence, $x_{1}$, $x_{2}$, and $x_{3}$ can be represented using the same number of binary variables. Quadratic terms, however, require connectivity between qubits, a limited computational resource in QAs.

To fully characterize the PBO, we must re-express each of the terms in $g_{p}$ and $g_{v}$ as products and sums of binary variables. However, it is convenient to simplify first their expressions. We start by substituting on the definition of $y'$ and collecting monomial terms:
\begin{equation*}
g_{p}(x_{1};\bar{x}_{2},\bar{x}_{3}) = \{-\ln(y) - \ln(1-x_{1}) - \beta \bar{x}_{2}- \beta (\mathbb{E}[\ln(z')|z] + \alpha \ln(y)) \bar{x}_{3} - \alpha \beta \ln(x_{1}) \bar{x}_{3} \}.
\label{eqn:simplified_gp_1}
\end{equation*}

Since $y$, $\bar{x}_{2}$, and $\bar{x}_{3}$ are constants in $g_{p}$, we can minimize the simplified version of the objective function:
\begin{equation*}
g_{p'}(x_{1};\bar{x}_{2},\bar{x}_{3}) = \{- \ln(1-x_{1}) - \alpha \beta \bar{x}_{3} \ln(x_{1}) \}.
\label{eqn:simplified_gp_2}
\end{equation*}
This reformulation will have additional benefits. Most importantly, it eliminates the $x_{2}$ term. We will see below why this elimination makes it easier to construct algorithms executed entirely on a QPU.

We can now write down a PBO for the policy improvement step. Since $a_{0}<0$, $\tilde{a}_{0} < 0$, and $\bar{x}_{3}>0$, we have $-(\tilde{a}_{0}+\alpha \beta a_{0} \bar{x}_{3}) > 0$. Consequently, removing the constant term might allow for negative energy states (loss function evaluations), which are problematic for one of the solution algorithms we will introduce in Section \ref{sec:Results}. Thus, we retain the constant and write:
\begin{equation}
\begin{split}
x_{1,0}^{*},...,x_{1,J_{1}}^{*} = & \underset{x_{1,0},...,x_{1,J_{1}}}{\text{argmin}} 
    \Bigg\{-(\tilde{a}_{0} + a_{0} \alpha \beta \bar{x}_{3}) - \sum_{j_{1}=0}^{J_{1}} \Bigg [(\tilde{a}_{1} + a_{1} \alpha \beta \bar{x}_{3}) 2^{j_{1}} \\
    & + a_{2} \alpha \beta \bar{x}_{3} 2^{2j_{1}} \Bigg ] x_{1, j_{1}} - 2 a_{2} \alpha \beta \bar{x}_{3} 
    \sum_{j_{1}=0}^{J_{1}} \sum_{i_{1} < j_{1}} 2^{j_{1} + i_{1}} x_{1, j_{1}} x_{1, i_{1}} \Bigg \}.
\label{eqn:gp_pbo}
\end{split}
\end{equation}

We next convert the policy valuation step into a PBO, starting with the expression:
\begin{equation*}
\begin{split}
g_{v}(x_{2},x_{3};\bar{x}_{1}) = \{\ln(y) + \ln(1-\bar{x}_{1}) - (1-\beta)x_{2} + \zeta x_{3} + \alpha \beta \ln(\bar{x}_{1}) x_{3} \}^{2},
\label{eqn:simplified_gv_1}
\end{split}
\end{equation*}
where we have defined $\zeta = \beta \mathbb{E}[\ln(z')|z] + (\alpha \beta - 1) \ln(y)$.

We now expand $g_{v}$:
\begin{equation}
\begin{split}
g_{v}(x_{2},x_{3};\bar{x}_{1}) = & \Bigg\{ \gamma_{0} + \gamma_{1}(\bar{x_{1}}) - \underbrace{2 (1-\beta)                                               (\ln(y)+\ln(1-\bar{x}_{1}))}_{\gamma_{2}(\bar{x}_{1})}  x_{2} \\
        & + \underbrace{ 2 \Bigg [\ln(y)(\zeta + \alpha \beta \ln(\bar{x}_{1})) + \ln(1-\bar{x}_{1})(\zeta + \alpha \beta \ln(\bar{x}_{1})) \Bigg ]}_{\gamma_{3}(\bar{x}_{1})} x_{3} \\
        & - \underbrace{2(1-\beta) \Bigg [\zeta  + \alpha \beta \ln(\bar{x}_{1}) \Bigg ]}_{\gamma_{23}(\bar{x}_{1})} x_{2} x_{3} + \underbrace{(1-\beta)^{2}}_{\gamma_{22}} x_{2}^{2} \\
        & + \underbrace{ \Bigg [\zeta^{2} + 2 \zeta \alpha \beta \ln(\bar{x}_{1}) + \alpha^{2} \beta^{2} \ln(\bar{x}_{1})^{2} \Bigg ]}_{\gamma_{33}(\bar{x}_{1})}  x_{3}^{2} \Bigg \}.
\label{eqn:simplified_gv_2}
\end{split}
\end{equation}

Notice that $\gamma_{0} = \ln(y)^{2} - 2(1-\beta)\ln(y)$ and $\gamma_{1}(\bar{x}_{1}) = 2\ln(y)\ln(1-\bar{x}_{1}) + \ln(1-\bar{x}_{1})^{2}$. We enclose $\bar{x}_{1}$ in parenthesis to indicate that the constant term $\gamma_{1}$ depends on $\bar{x}_{1}$ and will change after each policy improvement step.

Using the constants defined above, we rewrite equation \eqref{eqn:simplified_gv_2} more compactly as:
\begin{equation*}
g_{v}(x_{2},x_{3};\bar{x}_{1}) =   
    \Big\{\gamma_{0} + \gamma_{1}(\bar{x_{1}}) - \gamma_{2}(\bar{x}_{1}) x_{2} + \gamma_{3}(\bar{x}_{1}) x_{3} - \gamma_{23}(\bar{x}_{1}) x_{2} x_{3} + 
    \gamma_{22} x_{2}^{2} + \gamma_{33}(\bar{x}_{1}) x_{3}^{2} \Big \}.
\label{eqn:simplified_gv_3}
\end{equation*}

Finally, we define the PBO for the policy valuation step:
\begin{equation}
\begin{split}
x_{2,0}^{*},...,x_{2,J_{2}}^{*}, & x_{3,0}^{*},...,x_{3,J_{3}}^{*} = \underset{x_{2,0},...,x_{2,J_{2}}, x_{3,0},...,x_{3,J_{3}}}{\text{argmin}} 
    \Bigg\{\gamma_{0} + \gamma_{1}(\bar{x_{1}}) \\
    & - \gamma_{2}(\bar{x}_{1})  s_{2} \sum_{j_{2}=0}^{J_{2}} 2^{j_{2}} x_{2, j_{2}} 
    + \gamma_{3}(\bar{x}_{1}) s_{3} \sum_{j_{3}=0}^{J_{3}} 2^{j_{3}} x_{3, j_{3}} \\  
    & - \gamma_{23}(\bar{x}_{1}) s_{2}s_{3} \sum_{j_{2}=0}^{J_{2}} \sum_{j_{3}=0}^{J_{3}} 2^{j_{2}+j_{3}} x_{2, j_{2}} x_{3, j_{3}} \\
    & + \gamma_{22} s_{2}^{2} \left( \sum_{j_{2}=0}^{J_{2}} 2^{2*j_{2}} x_{2, j_{2}} + \sum_{j_{2}=0}^{J_{2}} \sum_{i_{2} < j_{2}} 2^{j_{2} + i_{2} + 1} x_{2, j_{2}} x_{2, i_{2}} \right) \\
    & + \gamma_{33}(\bar{x}_{1}) s_{3}^{2} \left( \sum_{j_{3}=0}^{J_{3}} 2^{2*j_{3}} x_{3, j_{3}} + \sum_{j_{3}=0}^{J_{3}} \sum_{i_{3} < j_{3}} 2^{j_{2} + i_{2} + 1} x_{3, j_{3}} x_{3, i_{3}} \right) \Bigg \}.
\label{eqn:gv_pbo}
\end{split}
\end{equation}

\subsection{Translating the PBF to a QUBO}
\label{ssec:methods_QUBO}

Converting a PBF into a QUBO entails quadratizing all cubic and higher-order terms. The policy improvement and valuation steps, as defined in equations \eqref{eqn:gp_pbo} and \eqref{eqn:gv_pbo}, do not contain any such terms. Consequently, no quadratization is needed if $x_{1}$ enters $g_{v}$ as a constant and $x_{2}$ and $x_{3}$ enter $g_{p}$ as a constant, as they do in the hybrid algorithm we propose in Section \ref{sec:Results}.

If, however, we want to reduce the computational overhead by solving the problem on the QPU entirely, as we also do in Section \ref{sec:Results}, then we will need to introduce higher-order terms. Subsection \ref{ssec:degree_reduction} explained how to quadratize such terms. We will also give two examples here in the context of our problem.

We first consider the case where all terms in $g_{p}$ are multiplied by the same binary variable, $x_{p}$. This will transform the quadratic terms into cubic terms in the last summation:
\begin{equation*}
     - \sum_{j_{1}=0}^{J_{1}} \sum_{i_{1} < j_{1}} 2 a_{2} \alpha \beta \bar{x}_{3} 2^{j_{1} + i_{1}} x_{1, j_{1}} x_{1, i_{1}} x_{p}.
    \label{eqn:quadratization_ex_1}
\end{equation*}

Since $a_{2} \alpha \beta \bar{x}_{3} 2^{j_{1}+i_{1}} > 0$ for all $\bar{x}_{3}$, $j_{1}$, and $i_{j}$, each term in the summation will be negative. Thus, we can use the NTR method described in Subsection \ref{ssec:degree_reduction} to quadratize the expression with a single auxiliary variable, $x_{a, i_{1}, j_{1}}$, for each term, yielding:
\begin{equation*}
     - \sum_{j_{1}=0}^{J_{1}} \sum_{i_{1} < j_{1}} 2 a_{2} \alpha \beta \bar{x}_{3} 2^{j_{1} + i_{1}} x_{a, i_{1}, j_{1}} (2 - x_{1, j_{1}} - x_{1, i_{1}} - x_{p})
    \label{eqn:quadratization_ex_2}
\end{equation*}
This step adds a total of $\frac{J_{1}(J_{1}-1)}{2}$ auxiliary variables to the policy valuation step.
 
Consider the case where $x_{1}$ does not enter $g_{v}$ as a constant. This happens whenever we use an algorithm solved exclusively on the QPU and, thus, cannot specify $x_{1}$ as a parameter. The term $\alpha^{2}\beta^{2} x_{3} x_{1}^{2}$, for example, becomes:
\begin{equation*}
\begin{split}
    \alpha^{2}\beta^{2} & x_{3}\ln(x_{1})^{2} = \alpha^{2}\beta^{2} s_{3} s_{1}^{2} \Bigg \{ a_{0}^{2} \sum_{j_{3}=0}^{J_{3}} 2^{j_{3}} x_{3, j_{3}}  + a_{0}(1+a_{1})\sum_{j_{1}=0}^{J_{1}} \sum_{i_{3}=0}^{J_{3}} 2^{j_{1}+i_{3}} x_{1,j_{1}} x_{3, j_{3}} \\
					& + (2a_{0}a_{2} + a_{1}^{2})\sum_{j_{1}=0}^{J_{1}} \sum_{i_{1}=0}^{J_{1}} \sum_{i_{3}=0}^{J_{3}} 2^{j_{1} + i_{1} + j_{3}} x_{1,j_{1}} x_{1,i_{1}} x_{3,j_{3}} + \\
					& + 2 a_{1} a_{2} \sum_{j_{1}=0}^{J_{1}} \sum_{i_{1}=0}^{J_{1}} \sum_{l_{1}=0}^{J_{1}} \sum_{j_{1}=0}^{J_{1}} 2^{j_{1} + i_{1} + l_{1} + j_{3}} x_{1,j_{1}} x_{1,i_{1}} x_{1,l_{1}} x_{3, j_{3}} \\
					& + a_{2}^{2} \sum_{j_{1}=0}^{J_{1}} \sum_{i_{1}=0}^{J_{1}} \sum_{l_{1}=0}^{J_{1}} \sum_{m_{1}=0}^{J_{1}} \sum_{j_{3}=0}^{J_{3}}	2^{j_{1} + i_{1} + l_{1} + m_{1} + j_{3}} x_{1, j_{1}} x_{1, i_{1}} x_{1, l_{1}} x_{1, m_{1}} x_{3, j_{3}} \Bigg \}.
		\label{eqn:x3x1sq}
\end{split}    
\end{equation*}
The final three summations all consist of cubic or higher-order terms and, thus, require quadratization. Additionally, since $s_{3} > 0$, all terms are positive, which precludes using NTR as a reduction method. We instead use the PTR method from \citet{BG14} and focus exclusively on the final summation in the expression, which consists of quintic terms. For each term, we will need three auxiliary variables:
\begin{equation*}
\begin{split}
    \alpha^{2}\beta^{2} & s_{3} s_{1}^{2} a_{2}^{2} \Bigg[ \sum_{j_{1}=0}^{J_{1}} \sum_{i_{1}=0}^{J_{1}} \sum_{l_{1}=0}^{J_{1}} \sum_{m_{1}=0}^{J_{1}} \sum_{j_{3}=0}^{J_{3}} x_{a_{1}, j_{1}, i_{1}, l_{1}, m_{1}, j_{3}} (3 + x_{1, j_{1}} - x_{1, i_{1}} - x_{1, l_{1}} \\
    & - x_{1, m_{1}} - x_{3, j_{3}}) + x_{a_{2}, j_{1}, i_{1}, l_{1}, m_{1}, j_{3}} (2 + x_{1, i_{1}} - x_{1, l_{1}} - x_{1, m_{1}} - x_{3, j_{3}}) + \\
    & x_{a_{3}, j_{1}, i_{1}, l_{1}, m_{1}, j_{3}} (1 + x_{1, l_{1}} - x_{1, m_{1}} - x_{3, j_{3}}) + x_{1, m_{1}} x_{3, j_{3}} \Bigg ].
		\label{eqn:x3x1sq_quadratization}
\end{split}    
\end{equation*}

We can also further reduce the number of auxiliary variables by exploiting symmetry and the properties of binary variables, as we have done previously.

\section{Results}\label{sec:Results}

So far, we have explained how to express our PPI problem as separate QUBO problems for the policy and value function updates. But we are still not quite ready to solve our problem on a QA because Algorithm \ref{alg:classical_ppi} alternates between policy and value function updates, an iterative structure QAs cannot handle.

To get around this problem, we proceed in four cumulative steps:

\begin{enumerate}
    
\item Subsection \ref{ssec:ClassicalAlg} runs the QUBO version of PPI, Algorithm \ref{alg:classical}, on a classical computer, iterating over value and policy function updates. This step allows us to compare the QUBO version of PPI with the original PPI in Algorithm \ref{alg:classical_ppi}, which we already ran on a classical computer (Figure \ref{fig:classical_solution}).

\item Subsection \ref{ssec:HybridAlg} proposes a new quantum-classical hybrid algorithm, Algorithm \ref{alg:hybrid}, where we run one part of the PPI on a classical computer and the rest on a QA. Using these hybrid algorithms has become popular in the literature \citep{HSDW20}. 

\item Subsection \ref{sssec:multi_anneal_quantum} considers a pure quantum algorithm, Algorithm \ref{alg:quantum}, that exploits reverse and inhomogeneous annealing to perform a complete iteration of policy and value function updates within an anneal. It also eliminates the need to reprogram the QPU and iterates across anneals.

\item Subsection \ref{sssec:oneshot_quantum} modifies Algorithm \ref{alg:quantum} to perform multiple iterations within a single anneal and delivers Algorithm \ref{alg:oneshot_quantum}, which yields a candidate solution from each anneal. Algorithms \ref{alg:quantum} and \ref{alg:oneshot_quantum} are the most innovative part of our paper from a technical perspective.

\end{enumerate}

Let us go over each step now in order.

\subsection{Classical Combinatorial PPI}
\label{ssec:ClassicalAlg}

First, we solve the PPI problem with a classical combinatorial algorithm that performs alternating iterations of policy improvement and valuation steps. This algorithm is the combinatorial version of Algorithm \ref{alg:classical_ppi}. The policy improvement step is solved analytically, whereas the policy valuation step is specified as a QUBO and solved exactly by searching over the discretized value function parameters. The details are given in Algorithm \ref{alg:classical}. 

\begin{algorithm}
\DontPrintSemicolon
Initialize the value function parameters, $x_{2}$ and $x_{3}$.\;
Compute $x_{1}$ analytically.\;
Given the QUBO and $x_{1}$, perform the policy valuation step by identifying the values of $\{ x_{2,0},...,x_{2,J_{2}} \}$ and $\{ x_{3,0},..,x_{3,J_{3}} \}$ that minimize the loss function.\;
Map the solution to value function parameters: $\{ x_{2,0},...,x_{2,J_{2}} \} \rightarrow x_{2}$ and $\{ x_{3,0},..,x_{3,J_{3}} \} \rightarrow x_{3}$.\;
Repeat steps 2-4 until the convergence criterion is satisfied.
\caption{Classical Combinatorial Parametric Policy Iteration}\label{alg:classical}
\end{algorithm}

Using an exact solution for $x_{1}$ provides us with a benchmark that has three advantages. First, it eliminates errors from the policy valuation step, yielding an approximate measure of errors arising from discretization. Second, it provides an upper bound for acceptable candidate solution run times since it is always better to use an exact solution when the execution time is lower than the alternatives. And third, it can be used to evaluate how many iterations will be needed to achieve convergence in the hybrid and quantum algorithms.

We parameterize the algorithm by setting $J_{2} = J_{3} = 9$, which allows $x_{2}$ and $x_{3}$ to each take on $2^{10}$ discrete values, and requires a search over $2^{20}$ states on each policy valuation step. We select scaling factors, $s_{2} = -0.035$ and $s_{3} = 0.003$, to bound the value function parameters, $x_{2} \in [0,2x_{2}^{*}]$ and $x_{3} \in [0,2x_{3}^{*}]$, where $x_{2}^{*}$ and $x_{3}^{*}$ correspond to the true parameter values.

\begin{figure}[!htb]
\includegraphics[width=\linewidth]{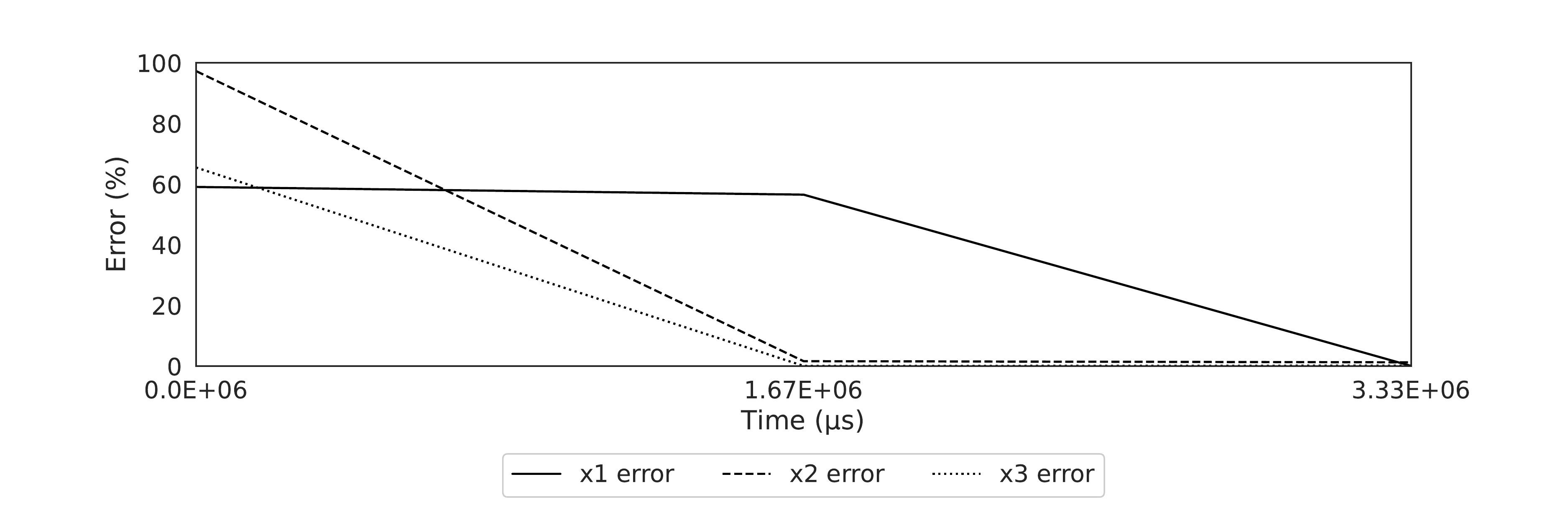}
\caption{Policy and value function parameter errors as a function of average execution time over 10 runs. Convergence is achieved after two iterations.}
\label{fig:classical_combinatorial_solution}
\end{figure}

Figure \ref{fig:classical_combinatorial_solution} plots the average results from 10 executions of the algorithm. The horizontal axis shows the total execution time in microseconds (only the run time varies across executions; the candidate solutions are identical). Each tick corresponds to an iteration. The vertical axis shows the parameter errors in absolute percentage deviations from their true values. The terminal errors for $x_{1}$, $x_{2}$, and $x_{3}$ are 0.04\%, 1.24\%, and 0.14\%, respectively. Other parameter configurations yield lower errors for individual parameters but at the expense of higher errors for other parameters and a higher loss. Convergence is achieved after 3.33 seconds (two iterations), which is slower than the VFI solution in \AFV{} and Algorithm \ref{alg:classical_ppi}. 

All run time comparisons here and below exclude initial pre-processing and terminal post-processing times. Since the benchmark we select is competitive and can be solved classically in under one second (see the results in \AFV{}), we restrict the comparisons to tasks where there are substantive differences across algorithms. Removing fixed and polynomial time computational overhead common to all algorithms simplifies this comparison.

The classical computational steps are performed on a Linux machine with an 8-core Intel Xeon 2.00GHz processor and 50GBs RAM. The classical algorithms and classical components of hybrid algorithms were programmed in Python. The quantum algorithms and quantum components of classical algorithms were executed on D-Wave's \textit{Advantage} System 6.1, which has 5616 qubits, arranged in a lattice of 16 x 16 titles known as a \textit{Pegasus} (P16) graph. We have released code that can be used to reproduce the results in the paper.\footnote{See the GitHub repository at \url{https://github.com/ijh85/quantum-dynamic-programming}, which contains code that can be used to verify the results in the paper.}

\subsection{Hybrid Quantum-Classical Algorithm}
\label{ssec:HybridAlg}

We next approach the PPI problem with a hybrid algorithm that performs alternating iterations of policy improvement and valuation steps. The policy valuation step is run on the QA, while the rest of the algorithm is run classically. The details are given in Algorithm \ref{alg:hybrid}. Notice that $[C]$ indicates a step is performed classically and $[Q]$ on a QA.

\begin{algorithm}
\DontPrintSemicolon
[C] Initialize the value function parameters, $x_{2}$ and $x_{3}$.\;
[C] Compute $x_{1}$ analytically.\;
[Q] Given the QUBO and $x_{1}$, perform the policy valuation step, yielding $\{ x_{2,0},...,x_{2,J_{2}} \}$ and $\{ x_{3,0},..,x_{3,J_{3}} \}$ for each anneal.\;
[C] Map the output of each anneal to value function parameters: $\{ x_{2,0},...,x_{2,J_{2}} \} \rightarrow x_{2}$ and $\{ x_{3,0},..,x_{3,J_{3}} \} \rightarrow x_{3}$.\;
[C] Compute the mean values of $x_{2}$ and $x_{3}$ for the lowest energy level anneals.\;
Repeat steps 2-5 until the convergence criterion is satisfied.
\caption{Hybrid Parametric Policy Iteration}\label{alg:hybrid}
\end{algorithm}

As before, we set $J_{2} = 9$, $J_{3} = 9$, $s_{2} = -0.035$, and $s_{3} = 0.003$, and use 100 anneals, each with a duration of $20 \mu s$ (microseconds). We then select the 10\% of anneals with the lowest energy levels and terminate the process after two iterations. The average energy level (loss) of the system tends to decline substantially over the first two iterations; however, subsequent steps have considerably smaller impacts on the loss. This aligns well with the classical results.

Notice that the hybrid algorithm contains quantum components that are affected by \textit{true} randomness that arises in quantum systems. Consequently, the candidate solutions that the algorithm produces cannot be reproduced from an initial state. For this reason, we focus on the distribution of candidate solutions produced by 50 executions of the algorithm.

\begin{table}[!ht]
\vspace{4mm}
\setlength{\tabcolsep}{5pt}
\begin{center}
\scalebox{0.95}{
\begin{tabular}{|l|c|c|c|c|c|c|}
\hline
& $x_{1}$ Error \% & $x_{2}$ Error \% & $x_{3}$ Error \% & QPU Total & QPU Prog. & Total Time \\
\hline 
Mean & 5.70 & 1.34 & 9.63 & 5.23E+04 & 3.19E+04 & 1.76E+06 \\
25th Percentile & 3.12 & 0.58 & 5.33 & 5.23E+04 & 3.19E+04 & 1.53E+06 \\
75th Percentile & 7.74 & 2.03 & 13.22 & 5.23E+04 & 3.19E+04 & 1.60E+06\\
Std. Dev. & 3.99 & 0.90	& 6.00 & 0.0E+04 & 0.00E+04 & 0.78E+06 \\
\hline
\end{tabular}}
\caption{The initial values are $x_{1} = 0.5$, $x_{2} = -0.5$, and $x_{3} = 0.5$. All times are given in microseconds. Errors for the policy and value function parameters are given as absolute percentage deviations from their true values. Summary statistics are reported from 50 executions of the same hybrid program. The quantum part of each iteration uses 100 anneals and selects the 10\% with the lowest energy levels.}
\label{tab:hybrid_results}
\end{center}
\end{table}

Table \ref{tab:hybrid_results} reports the results. The error columns for $x_{1}$, $x_{2}$, and $x_{3}$ provide the absolute percentage deviation of the initial parameter values from their true values. The subsequent three columns show the total amount of computation time on the quantum processor (QPU Total), the amount of time needed to program the quantum processor (QPU Prog.), and the total execution time for both the classical and the quantum components (Total Time).

The hybrid algorithm tends to converge to a neighborhood of $\{x_{1}^{*}$, $x_{2}^{*}$, $x_{3}^{*} \}$ in two iterations. On average, it takes 1.76 seconds, 1.9 times faster than the classical combinatorial solution but 2.4 times slower than the fastest \texttt{C++} implementation for VFI in \AFV{}, which converged in 0.73 seconds. Importantly, only 3\% of the hybrid algorithm's execution time comes from QPU computation. And only 39\% of that is attributable to annealing. The rest is overhead from the QPU programming step, a computational cost that precedes a sequence of anneals for a given problem.

Thus, moving to a non-hybrid quantum algorithm could substantially improve computational time. An algorithm requiring the QPU to be programmed only once could reduce the total execution time from 1.76 seconds to 0.026 seconds.

Finally, notice that the hybrid algorithm yields less precise results on average than the classical benchmarks. The terminal value of $x_{3}$, for instance, deviates from $x_{3}^{*}$ by 9.63\% on average. This is in line with the literature, which finds that QAs can produce high-quality approximations to the global solution fast but require a substantial amount of time to refine the solution further \citep{ZBT+15, KYN+15}. If higher precision is needed, the QA solution can be used to warm-start a classical algorithm.

\subsection{Multi-Anneal Quantum Algorithm}
\label{sssec:multi_anneal_quantum}

Motivated by the previous results, we construct an algorithm that eliminates the classical components of the algorithm and requires the QPU to be programmed only once. More concretely, we specify a single QUBO, along with an annealing schedule, that can iteratively optimize the two components of the objective function. To so do, we introduce two novel computational strategies, both of which use recent developments: reverse annealing and inhomogeneous annealing. 

Reverse annealing initializes the QPU in a classical state rather than an equal superposition over the entire state space. It then opens up a superposition state weighted toward the initial classical state. Specifying a full reversal is equivalent to performing a standard forward anneal, whereas specifying a small, partial reversal is likely to yield a solution close to the initial classical state. In standard usage, reverse annealing is applied to refine candidate solutions by searching within a specified neighborhood \citep{PHD20}. The extent of the reversal determines the size of the neighborhood.

Inhomogeneous annealing supplies the QPU with separate offset values for each qubit. A positive offset indicates that the specified qubit should be annealed ahead of the global schedule, while a negative value indicates that it should be annealed afterward. Inhomogeneous annealing has been applied to improve alignment between multiple qubits in the annealing process \citep{AM20}.

We use reverse and inhomogeneous annealing to facilitate iteration over two separate objective functions, which differs from standard practice. Algorithm \ref{alg:quantum} performs this iteration across anneals. Algorithm \ref{alg:oneshot_quantum}, which we will introduce in the next subsection, demonstrates how to iterate within an anneal. In both cases, the QPU is only programmed once.

Algorithm \ref{alg:quantum} describes the multi-anneal quantum routine. We parameterize the algorithm by setting $J_{1} = J_{2} = J_{3} = 6$, which allows $x_{1}$, $x_{2}$, and $x_{3}$ to each take on $2^{7}$ discrete values. 

\begin{algorithm}
\DontPrintSemicolon
Quadratize the policy improvement and valuation PBFs separately, reconcile the resulting QUBOs, and then merge them.\;
Initialize the classical states of policy function variables, $\{x_{1,0},...,x_{1,J_{1}}\}$, and value function variables, $\{x_{2,0},...,x_{2,J_{2}}, x_{3,0},...,x_{3,J_{3}}\}$.\;
Initialize variables that activate and deactivate loss components, $x_{p} = 0$ and $x_{v} = 0$.\;
Set an inhomogeneous annealing schedule that first anneals $\{x_{1,0},...,x_{1,J_{1}}\}$ and $x_{p}$, and then anneals $\{x_{2,0},...,x_{2,J_{2}}, x_{3,0},...,x_{3,J_{3}}\}$ and $x_{v}$.\;
Set a reverse annealing schedule that applies to all qubits.\;
Perform $\mathcal{N}$ anneals without reinitializing the original classical state.\;
Select the results from the anneal with the lowest energy level.
\caption{Quantum Parametric Policy Iteration}\label{alg:quantum}
\end{algorithm}

Step 1 uses a non-standard approach to quadratize the PBF into a QUBO. This entails quadratizing $x_{p}g_{p}$ and $x_{v}g_{v}$ separately, reconciling the two resulting QUBOs, and merging them. This approach ensures that qubits are otherwise inactive and annealed with the correct group. Steps 2 and 3 initialize the QA in a classical state where both components of the objective function are inactive (i.e., $x_{p}=0$ and $x_{v}=0$). In steps 4 and 5, the reverse anneal opens up a superposition for $\{x_{1,0},...,x_{1,J_{1}}\}$ and $x_{p}$, allowing for optimization over the policy function and with the policy function parameters, but not over the value function or with the value function parameters. As the anneal progresses through its schedule, the qubit offsets specify that a superposition should be opened for $\{x_{2,0},...,x_{2,J_{2}}, x_{3,0},...,x_{3,J_{3}}\}$ and $x_{v}$, allowing for optimization over the value function and with value function parameters. The modified objective function is given by:
\begin{equation*}
g = x_{p} g_{p}(x_{1};\bar{x}_{2},\bar{x}_{3}) + x_{v} g_{v}(x_{2},x_{3};\bar{x}_{1})
\label{eqn:gq_objective}
\end{equation*}

Two additional details complete the algorithm. First, the QA should not be set to reinitialize after each anneal. This will ensure that each anneal in a sequence starts in the terminal classical state of the previous anneal, allowing for the repetition of both the policy improvement and the valuation steps. Second, the problem and annealing schedule should reliably yield $x_{p}=0$ and $x_{v}=0$ as the terminal classical states. This is because $x_{p}$ must converge to $0$ to ensure that the policy function component of the objective is inactive (or small) during the policy valuation step. It also ensures that the system starts the next anneal with $x_{p}=0$ and $x_{v}=0$. This can always be achieved by adding a positive bias term to one of the components of the objective function. The term's magnitude can be increased until anneals reliably yield $x_{p}=0$ and $x_{v}=0$.

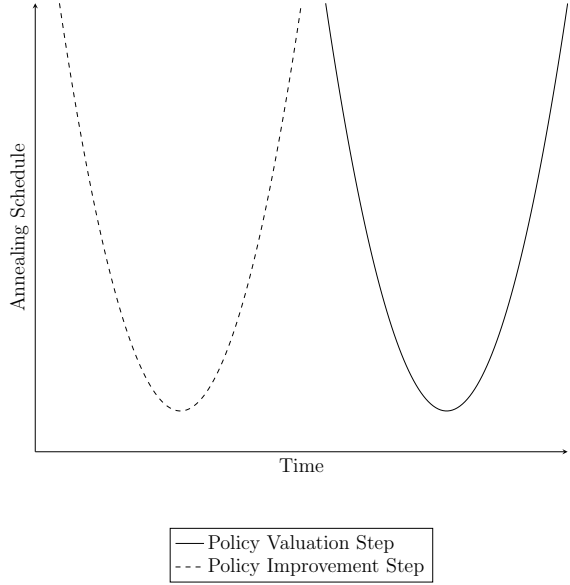
\begin{figure}[htb!]
\begin{center}
\resizebox{0.45\textwidth}{!}{
\begin{tikzpicture}
\begin{axis}[
    ymin = 0,
    ymax = 1.1,
    xmin = 0,
    ticks = none,
    axis lines = left,
    xlabel = {Time},
    ylabel = {Annealing Schedule},
    legend style={at={(0.5,-0.17)},anchor=north,legend cell align=left},
]
\addplot [
    domain=12:22, 
    samples=100, 
    color=black,
    ]
    {0.10 + 0.04*(x-17)^2};
\addplot [
    domain=1:11, 
    samples=100,
    color=black,
    style=dashed,
    ]
    {0.10 + 0.04*(x-6)^2};
\addlegendentry{Policy Valuation Step}
\addlegendentry{Policy Improvement Step}
\end{axis}
\end{tikzpicture}
}
\end{center}
\caption{Stylized depiction of the annealing schedule.}
\label{fig:annealing_schedule}
\end{figure}

\begin{figure}[!htb]
\minipage{0.45\textwidth}
  \includegraphics[width=\linewidth]{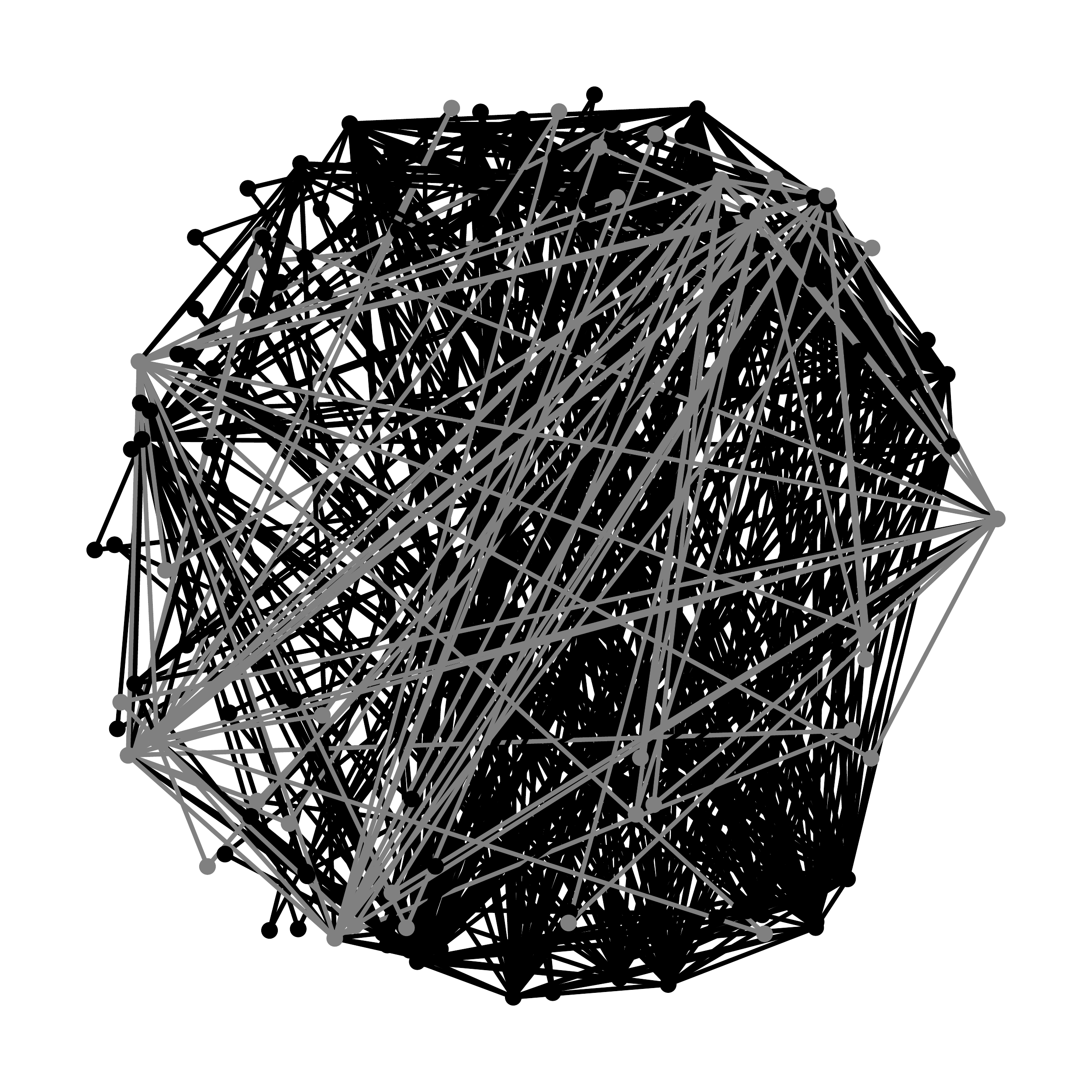}
  \caption*{Policy Improvement Graph}
\endminipage\hfill
\minipage{0.45\textwidth}
  \includegraphics[width=\linewidth]{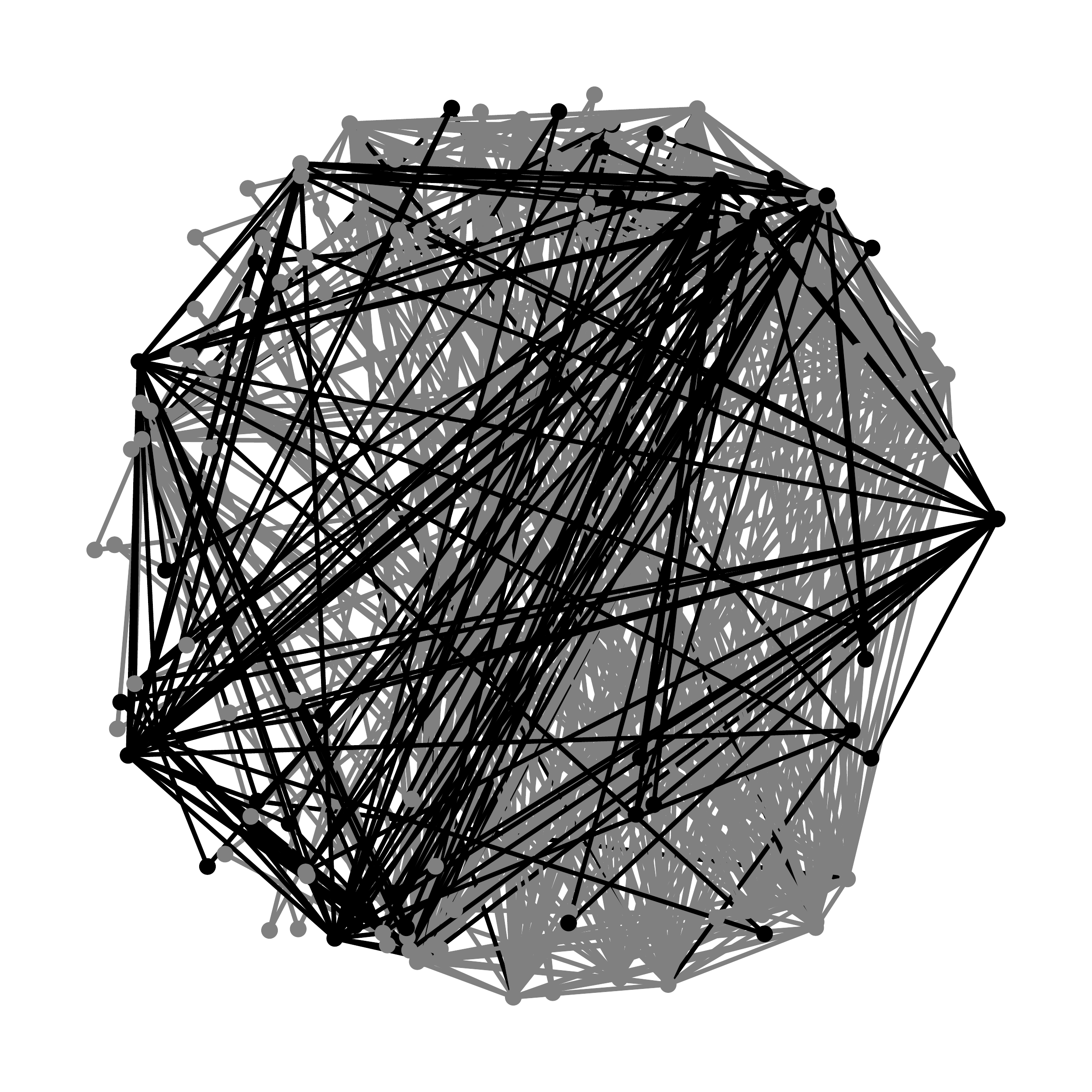}
  \caption*{Policy Valuation Graph}
\endminipage
\caption{State of the problem graph during the policy improvement and policy valuation steps.}
\label{fig:qpu_solution}
\end{figure}

Figure \ref{fig:annealing_schedule} provides a stylized depiction of the annealing schedule. The policy function function parameters and the objective function component are annealed first (dashed line), followed by the value function and its parameters (continuous line). In both cases, the parameters start in the classical state and are then reverse annealed and forward annealed. The annealing activity in each step is depicted as a weighted graph in Figure \ref{fig:qpu_solution}. A gray node indicates that a qubit is in a classical state, whereas a black node indicates that it is a superposition state undergoing an anneal. Black edges indicate that at least one of the connected nodes is being annealed. Notice that the policy improvement step has more annealing activity, since the quadratic terms used to approximate $\ln(x_{1})$ require connectivity.

Table \ref{tab:qpu_baseline_results} provides summary statistics for 50 executions of Algorithm \ref{alg:quantum}, each performing 20 anneals. For each QPU execution, we select the parameter values associated with the lowest-energy anneals for each objective function.\footnote{The terminal energy levels of the annealer are not a reliable metric for comparing anneals, since we use a non-standard objective function that is zeroed out over the anneal schedule. Instead, we reconstruct the two components of the objective function, as described in Subsection \ref{sssec:multi_anneal_quantum}.} The total QPU computation time is now reduced to 0.0217 seconds, approximately 3\% of the 0.73 second computation time of the VFI solution using \texttt{C++} in \AFV{} or 0.66\% of the computation time of the classical combinatorial algorithm in Subsection \ref{ssec:ClassicalAlg}.

\begin{table}[!ht]
\vspace{4mm}
\setlength{\tabcolsep}{5pt}
\begin{center}
\scalebox{0.9}{
\begin{tabular}{|l|c|c|c|c|c|c|}
\hline
 & $x_{1}$ Error \% & $x_{2}$ Error \% & $x_{3}$ Error \% & QPU Total & QPU Programming Time \\
\hline 
Mean & 12.13 & 5.82 & 14.50 & 2.17E+04 & 1.60E+04 \\
25th Percentile & 9.62 & 3.33 & 7.40 & 2.17E+04 & 1.60E+04 \\
75th Percentile & 14.82 & 7.21 & 20.22 & 2.17E+04 & 1.60E+04 \\
Standard Deviation & 3.58 & 3.62 & 9.26 & 0.00E+00 & 0.00E+00 \\
\hline
\end{tabular}}
\caption{We use a reverse annealing parameter of 0.00, which corresponds to a 100\% reversal. All times are given in microseconds. Summary statistics are reported from 50 QPU executions of the same program, each of which performs 50 anneals.}
\label{tab:qpu_baseline_results}
\end{center}
\end{table}

While the run time is reduced, the mean policy and value function parameter errors across the 50 QPU executions increase slightly relative to the hybrid algorithm. Also, iterating across anneals without resetting the QPU results in high correlation across candidate solutions within a QPU execution. Consequently, we execute the QPU multiple times to get independent solution candidates. In the next subsection, we propose a refinement that transforms each anneal into an independent candidate solution. This allows us to reliably obtain high-quality results from programming and executing the QPU once.

\subsection{One-Shot Quantum Algorithm}
\label{sssec:oneshot_quantum}

Next, we propose a \textit{one-shot} quantum PPI algorithm, Algorithm \ref{alg:oneshot_quantum}. In contrast to Algorithm \ref{alg:quantum}, Algorithm \ref{alg:oneshot_quantum} allows for multiple iterations over the policy and value functions within a single anneal. In addition, it produces terminal energy levels that can be directly compared across anneals, allowing us to reliably select the candidate solutions with the lowest associated losses. This is not possible with Algorithm \ref{alg:quantum} because $x_{p}$ and $x_{v}$ converge to zero over the annealing step. Relative to Algorithm \ref{alg:quantum}, Algorithm \ref{alg:oneshot_quantum} has two drawbacks: it requires longer annealing schedules and longer pauses between anneals to reinitialize the state.

\begin{algorithm}
\DontPrintSemicolon
Quadratize the policy improvement and valuation PBFs separately, reconcile the resulting QUBOs, and then merge them.\;
Initialize random classical states for the policy and value function variables: $\{x_{1,0},...,x_{1,J_{1}},x_{2,0},...,x_{2,J_{2}}, x_{3,0},...,x_{3,J_{3}}\}$.\;
Initialize (de-)activation variables in classical zero states: $x_{v} = x_{p} = 0$.\;
Set an inhomogeneous annealing schedule to activate the policy improvement ($g_{p}$) and policy valuation ($g_{v}$) functions in sequence, and to anneal their associated variables $\mathcal{C}$ times.\;
Set the global annealing schedule to perform a full reversal.\;
Perform $\mathcal{N}$ anneals and reinitialize the state after each.\;
Compute alternative measures of terminal state energy for each anneal.\;
Return average parameter values over the lowest energy anneals.\;
\caption{One-Shot Quantum Parametric Policy Iteration}\label{alg:oneshot_quantum}
\end{algorithm}

By allowing for iteration within the annealing step, we can perform a full reverse anneal and search the entire state space, rather than searching just a neighborhood around the terminal state of the previous anneal. Consequently, the terminal state of each anneal may be treated as a candidate solution rather than an iteration. Addtionally, since no information needs to be retained across anneals, we may reinitialize the state to reduce the correlation across candidate solutions.

Algorithm \ref{alg:oneshot_quantum} contains two novel components: 1) an inhomogeneous reverse annealing schedule that permits multiple iterations \textit{within} an anneal; and 2) a post-processing routine that boosts the quality of solutions. Let us discuss each of them before executing the algorithm and discussing the results.

\paragraph{Annealing schedule.}

\begin{figure}[!htb]
\begin{subfigure}{\textwidth}
    \includegraphics[width=0.95\linewidth]{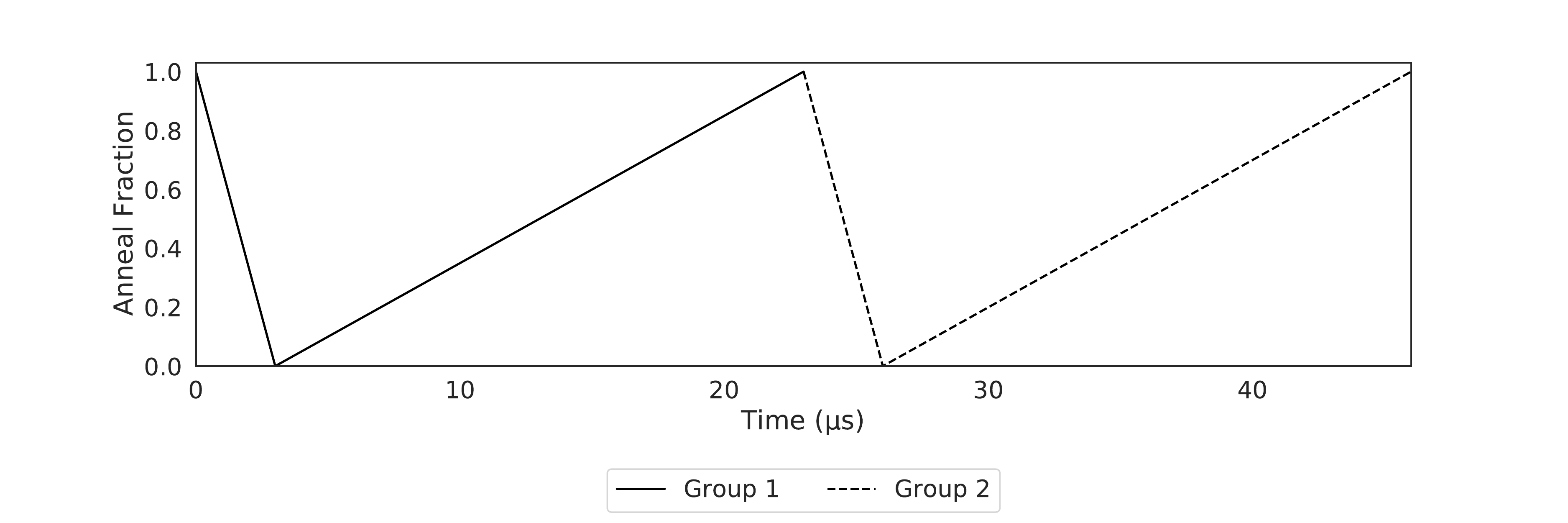}
\caption*{One Cycle Experimental Annealing Schedule}
\end{subfigure}
\begin{subfigure}{\textwidth}
    \includegraphics[width=0.95\linewidth]{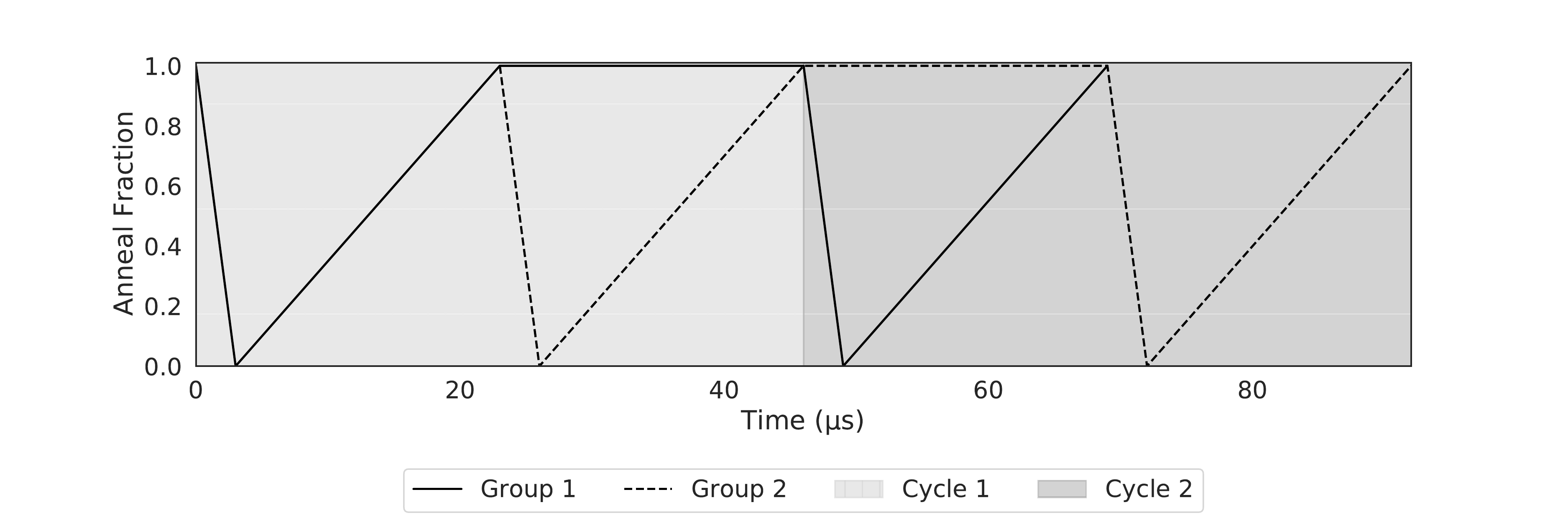}
\caption*{Two Cycle Experimental Annealing Schedule}
\end{subfigure}
\begin{subfigure}{\textwidth}
    \includegraphics[width=0.95\linewidth]{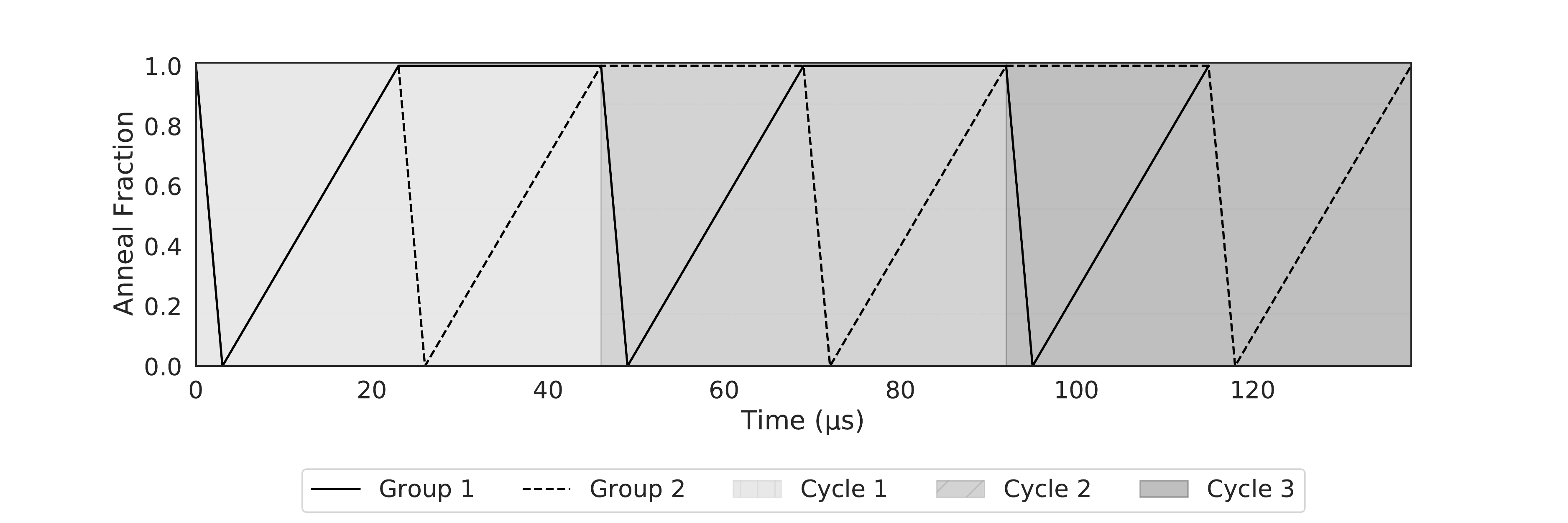}
\caption*{Three Cycle Experimental Annealing Schedule}
\end{subfigure}
\caption{Experimental annealing schedules for the case of 1, 2, and 3 cycles. In all cases, we adopt a reverse, inhomogeneous anneal. Group 1 contains qubits $\{z_{0}\}$ for $\mathcal{H}_{s}$ and $\{z_{0},z_{2}\}$ for $\mathcal{H}_{c}$. Group 2 contains qubits $\{z_{1}\}$ for $\mathcal{H}_{s}$ and $\{z_{1},z_{3}\}$ for $\mathcal{H}_{c}$. All times are given in microseconds ($\mu s$).}
\label{fig:experiment_anneal_schedule}
\end{figure}

Figure \ref{fig:experiment_anneal_schedule} illustrates the annealing schedule we propose for the one-shot algorithm, which contains reversals and is inhomogeneous.

The top panel illustrates the case where qubits are partitioned into two groups annealed separately but within a single \textit{cycle}. We will use the term cycle to refer to a subschedule where all qubits are annealed once. The top panel contains one such subschedule and, thus, one cycle, whereas the middle and bottom panels contain two and three cycles since they are repetitions of the subschedule in the top panel.

Our proposed algorithm deviates from the original conception of quantum annealing, which was structured around a simple forward anneal \citep{FGGS00}. Furthermore, existing applications of reverse annealing typically focus on solution refinement \citep{PHD20}, whereas we exploit reverse annealing to construct iterative algorithms. To demonstrate that our Algorithm \ref{alg:oneshot_quantum} performs as intended, we conduct two experiments in a simpler setting and report the results in Online Appendix \ref{sec:annealing_schedule_simulations}.

As demonstrated in the simulations, cycling over repetitions of the same annealing subschedule allows us to emulate an iterative algorithm within a single anneal. This suggests that the annealing schedule in Algorithm \ref{alg:oneshot_quantum}, coupled with the (de-)activation mechanism, could provide the foundation for a one-shot quantum algorithm.

Finally, it is worth explaining why the initialization matters, since we use a full anneal reversal. As shown in Figure \ref{fig:experiment_anneal_schedule}, the qubits are partitioned into two groups, which are annealed separately in each cycle. As such, one group will be in a classical state while the other is reversed into a full superposition state prior to the forward annealing step. The initialization aims to pin down the classical states of Group 2 qubits during the first round of annealing for Group 1. However, Group 1 qubits are immediately reversed into a uniform superposition and, thus, may be initialized in any state.

\paragraph{Post-processing.}

Algorithm \ref{alg:oneshot_quantum} yields a candidate solution from each anneal. However, there are often substantial differences in the quality of the candidate solution. This is captured by the system's terminal energy level, which is used in more standard problems to identify the best candidate solution. As discussed in conjunction with Algorithm \ref{alg:quantum}, these energy level readouts are not a reliable measure of solution quality for our problem since the weights of different components of the Hamiltonian are shrunk during parts of the annealing schedule.

Thus, we modify the post-processing routine to improve the interpretability of energy levels. We start by recomputing the energy level in each terminal state without applying the de-activation qubits. We refer to this measure as the \textit{unadjusted loss} since it provides an accurate measure of the loss but makes no further transformations. As we will see, however, this measure of the loss is still imperfect since it combines losses from both objective functions and is dominated by errors in $x_{2}$. In practice, the unadjusted loss will be helpful for evaluating the quality of a solution for $x_{2}$, but not for $x_{1}$ and $x_{3}$.

To account for these issues, we construct another measure of loss, which we refer to as the \textit{minimum loss}. This measure isolates the impact of errors in parameter $x_{j}$ on the component of the loss function associated with $x_{j}$. It does this by fixing the values of the other parameters at their true values. For example, for the policy function parameter, we would compute the minimum loss associated with a candidate solution, $x_{1}^{c}$, as $g_{p}(x^{c}_{1};x_{2}^{*},x_{3}^{*})$. This accomplishes two things. First, it reduces the variance in energy levels across anneals attributable to errors in the other parameters. And second, it ensures that the correct value of the parameter of interest (e.g., $x_{1}^{*}$) will be consistent with the lowest possible energy level. 

Unfortunately, it is not generally feasible to compute the minimum loss, since we would not typically know the true values of other parameters. Thus, we will use instead the \textit{adjusted loss}, which is feasible to construct and approximates the minimum loss. We compute this measure by dropping the $x_{2}$ terms from $x_{p}$, where they only enter additively. We then compute the loss separately for each parameter, fixing the values of all other parameters at their mean values over the lowest energy anneal solutions. This eliminates much of the variance across anneals in $g_{p}$ that is attributable to errors in $x_{2}$ and $x_{3}$.

\begin{table}[!ht]
\vspace{4mm}
\setlength{\tabcolsep}{5pt}
\begin{center}
\scalebox{1.00}{
\begin{tabular}{|c|c|c|c|}
\hline
& $x_{1}$ Error \% & $x_{2}$ Error \% & $x_{3}$ Error \% \\
\hline 
Unadjusted Loss & -0.01	& 0.92 & 0.05 \\
Minimum Loss & 0.97	& 0.96 & 0.97 \\
Adjusted Loss & 0.97 & 0.96 & 0.97 \\
\hline
\end{tabular}}
\caption{Correlations between the percentage error in the solutions for the policy and value function parameters.}
\label{tab:postselection_correlations}
\end{center}
\end{table}

As a final step, we retain the subset of anneals with the lowest adjusted loss for a given parameter and compute the mean value of the parameter over those anneals. Table \ref{tab:postselection_correlations} compares the correlations among parameter errors and the values of the three losses above. As discussed previously, there is a strong relationship between the unadjusted loss and the error in $x_{2}$, almost as strong as the minimum loss. In comparison, the correlations between the errors in $x_{1}$ and $x_{3}$ and the unadjusted loss is weak. This suggests that the unadjusted loss is not a helpful metric. The minimum loss, however, is strongly correlated with the parameter errors, suggesting that the elimination of errors in $x_{2}$ and $x_{3}$, for example, would enable us to identify anneals that contain useful information about $x_{1}^{*}$. The adjusted loss appears to be largely sufficient for this task, yielding a correlation coefficient of 0.97 for $x_{1}$, 0.96 for $x_{2}$, and 0.97 for $x_{3}$.

\begin{figure}[!ht]
\begin{subfigure}{0.9\textwidth}
    \centering
    \includegraphics[width=0.9\linewidth]{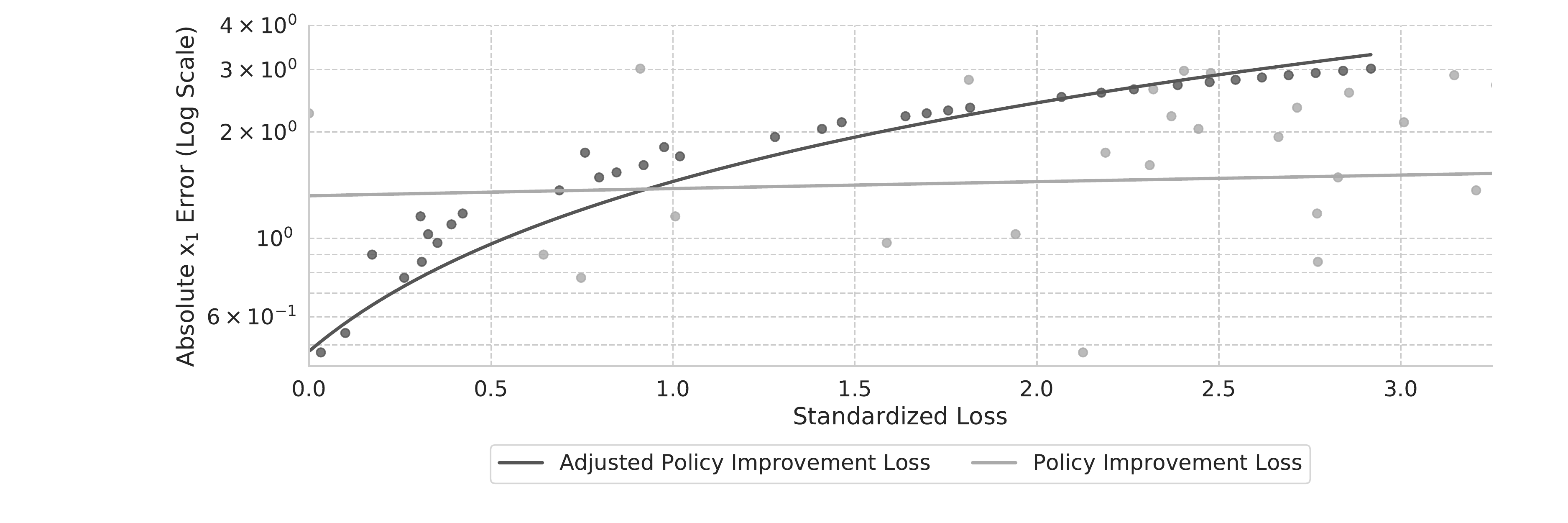}
    \caption*{$x_{1}$ Error and Policy Improvement Loss}
\end{subfigure}
\vspace{3mm}
\begin{subfigure}{0.9\textwidth}
    \centering
     \includegraphics[width=0.9\linewidth]{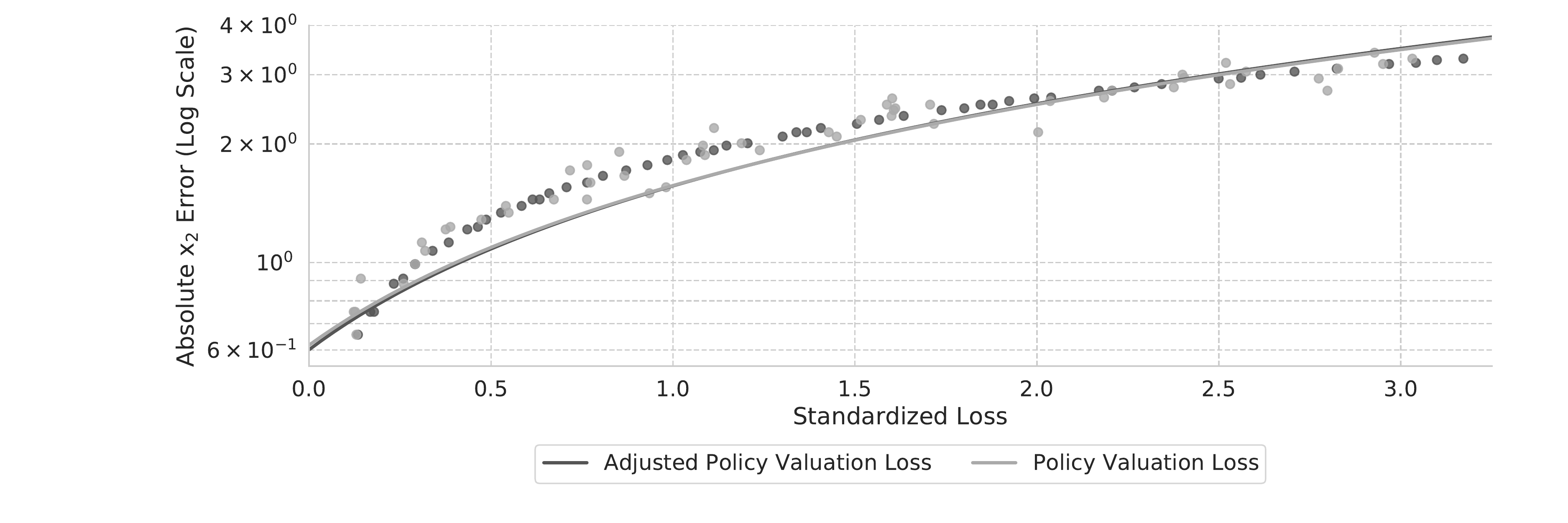}
    \caption*{$x_{2}$ Error and Policy Valuation Loss}
\end{subfigure}
\vspace{3mm}
\begin{subfigure}{0.9\textwidth}
    \centering
    \includegraphics[width=0.9\linewidth]{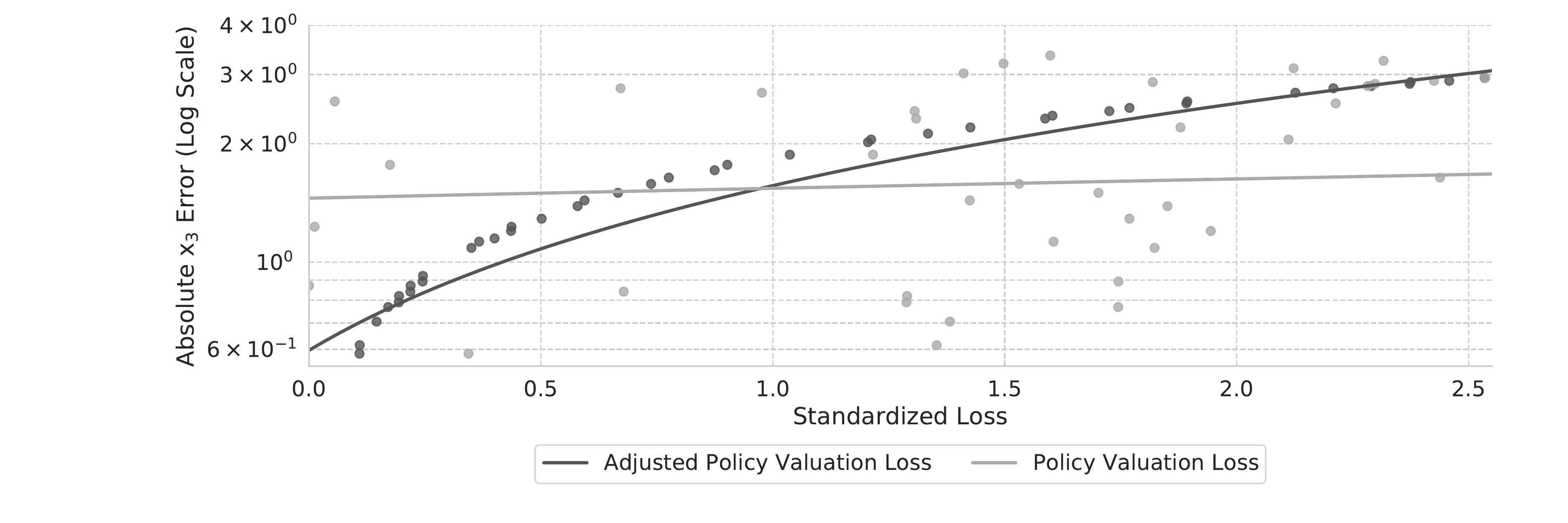}
    \caption*{$x_{3}$ Error and Policy Valuation Loss}
\end{subfigure}
\caption{Parameter error in absolute percentage deviations with levels of the \textit{unadjusted loss} and \textit{adjusted loss}. This is visualized using a binned scatter plot and fitted line for the results from 200 anneals. The vertical axis uses a log scale because the absolute error typically increases sharply at low levels of the \textit{adjusted loss}.}
\label{fig:postselection_figures}
\end{figure}

Figure \ref{fig:postselection_figures} further illustrates the benefits of post-processing anneals using the adjusted loss rather than the unadjusted loss. Each subfigure shows a binned scatterplot of loss values against absolute parameter value errors. The adjusted and unadjusted losses are roughly equivalent for $x_{2}$. However, for $x_{1}$ and $x_{3}$, the low values of the adjusted loss are highly informative about true parameter values, while the low values of the unadjusted loss are much less so.

\paragraph{Results.}

Table \ref{tab:qpu_oneshot_results} reports descriptive statistics for 50 QPU executions of Algorithm \ref{alg:oneshot_quantum}. In each case, we perform 200 anneals. Within each anneal, we execute three cycles of the annealing subschedule, as illustrated in the third panel of of Figure \ref{tab:qpu_oneshot_results}. The errors for the policy and value functions are lower on average than those generated by the multi-anneal quantum algorithm. Also, the errors for $x_{1}$ and $x_{3}$ are lower on average than those generated by the hybrid algorithm. The total execution time on the QPU is 0.065 seconds, which is 3.8 times as long as the multi-anneal quantum algorithm, but 3.7\% of the run time of the hybrid algorithm and 8.9\% of the run time of the \texttt{C++} implementation of VFI in \AFV{}, which was 0.73 seconds.

\begin{table}[!ht]
\vspace{4mm}
\setlength{\tabcolsep}{5pt}
\begin{center}
\scalebox{0.9}{
\begin{tabular}{|l|c|c|c|c|c|c|}
\hline
 & $x_{1}$ Error \% & $x_{2}$ Error \% & $x_{3}$ Error \% & QPU Total & QPU Programming Time \\
\hline 
Mean & 2.98 & 1.71 & 4.52 & 6.52E+04 & 1.59E+04 \\
25th Percentile & 1.70 & 1.08 & 1.92 & 6.52E+04 & 1.59E+04\\
75th Percentile & 2.49 & 2.01 & 3.98 & 6.52E+04 & 1.59E+04 \\
Standard Deviation & 3.48 & 0.87 & 5.47 & 0.00E+04 & 0.00E+04 \\
\hline
\end{tabular}}
\caption{We use a reverse annealing parameter of 0.0, which corresponds to a 100\% reversal. All times are given in microseconds. All anneals use three cycles. Summary statistics are reported from 50 QPU executions of the same program, each of which performs 200 anneals.}
\label{tab:qpu_oneshot_results}
\end{center}
\end{table}

The run time increases relative to the multi-anneal algorithm for three reasons. First, the annealing schedule contains multiple cycles of reverse and forward annealing, which increases the time per anneal from 23$\mu s$ to 115$\mu s$. Second, the classical state needs to be reinitialized between anneals to increase the independence of candidate results, adding overhead to each anneal. And finally, we use 200 anneals to identify independent candidate solutions, rather than iterating over a few highly correlated anneals.

\begin{figure}[!htb]
\begin{center}
\includegraphics[width=0.90\linewidth]{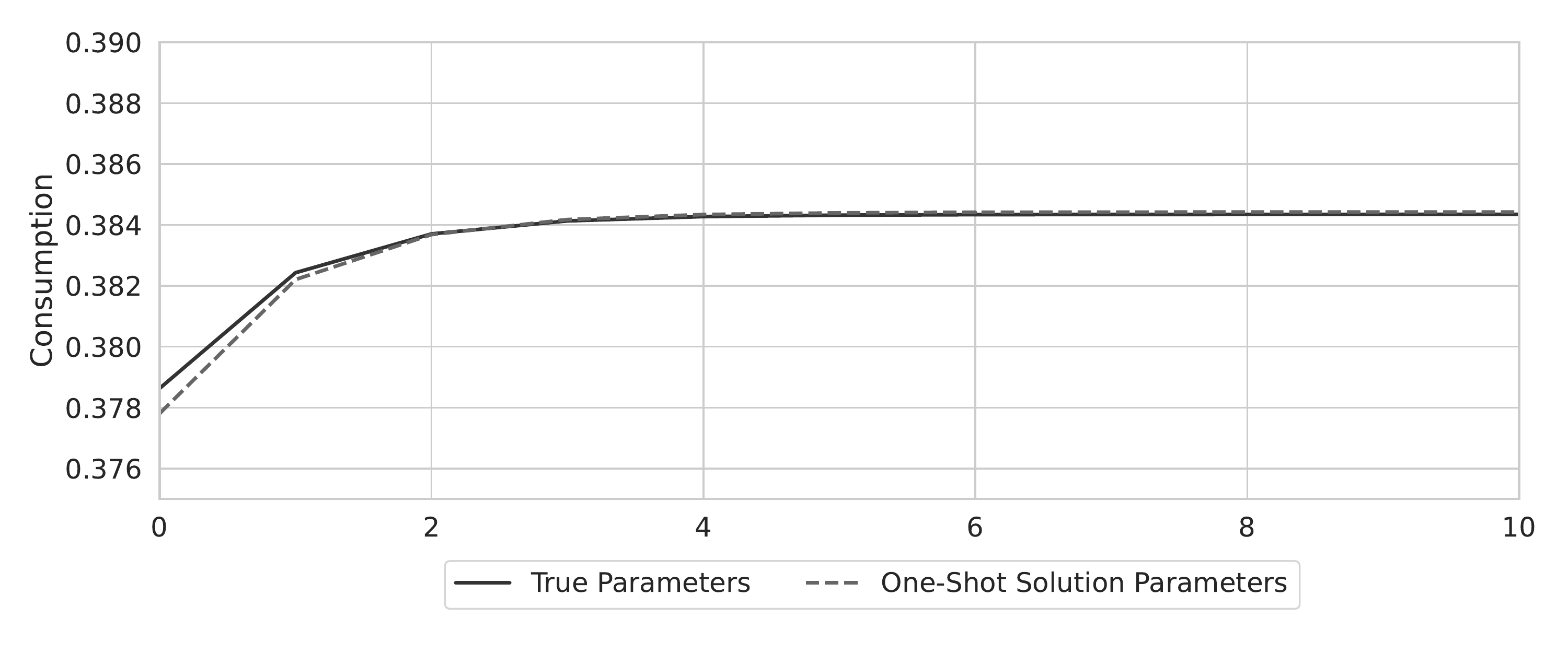}
\caption{Consumption simulation paths generated from the true parameters and the one-shot solution parameters. In both cases, we simulate the 10-period consumption response to a negative productivity shock in the first period.}
\label{fig:simulation}
\end{center}
\end{figure}

To provide further intuition for the size of the errors, we conduct a 10-period simulation exercise in which we compute the consumption path following a negative productivity shock. Figure \ref{fig:simulation} shows the paths implied by the analytical solution and the one-shot quantum algorithm solution. The difference is negligible in each period but is most pronounced immediately after the shock.

In summary: Algorithm \ref{alg:oneshot_quantum} provides a quantum solution that fully explores the state space, reduces correlation across the terminal states, and generates higher-quality solutions than the multi-anneal algorithm. It also provides an order of magnitude speed-up over classical benchmarks but is slower than the multi-anneal algorithm.

\section{Conclusion}
\label{sec:Discussion}

We have shown how to solve dynamic programming problems on QAs and constructed novel hybrid-classical and pure quantum algorithms that provide speed-ups over existing methods. We have illustrated our new algorithms by computing the RBC model in \AFV{}. We have also introduced a method for performing dynamic programming on QAs that has not previously been considered in the quantum computing literature. We have achieved run times that approach the theoretical minimum for our computational task on a QA and are an order of magnitude lower than the solutions explored in \AFV{} for the same RBC model. However, we see some reduction in accuracy, which is consistent with the existing literature on QA.

The choice of the RBC model does, however, rule out demonstrating quantum supremacy. With existing technology, the minimum time needed to program the quantum processing unit in a QA is nine milliseconds. This fixed computational cost will limit the achievable speed-up over the \texttt{C++} and Fortran solutions from \AFV{}. Demonstrating quantum supremacy would require us to use a much more complex model and one beyond what existing QAs can handle.

We have also discussed other uses of quantum annealing for dynamic programming. Quantum annealing could, for instance, be used to warm-start high-dimensional dynamic programming problems by mapping the parametric solution produced by the QA to initial values for the dynamic programming table employed by a more familiar classical algorithm.

Quantum annealing can also identify the average or maximum error size for a dynamic programming problem on a large grid. This could be useful if we have a candidate solution but cannot evaluate its accuracy in regions far from the steady state without significant computational expense. It could also identify features of the policy or value function that lead to performance degradation if not accounted for explicitly.

In summary, we hope to have convinced the reader that quantum computing opens many new doors in solving economic models. QAs are already good enough to solve some problems of interest in economics faster than even the fastest solutions using classical hardware. Much research, though, remains to be done.

\clearpage

\bibliography{References}

\clearpage

\appendix
\renewcommand\thepage{A.\arabic{page}}
\setcounter{page}{1}
\setcounter{figure}{0}
\renewcommand\thefigure{\thesection.\arabic{figure}}
\setcounter{table}{0}
\renewcommand\thetable{\thesection.\arabic{table}}
\setcounter{equation}{0}
\renewcommand\theequation{\thesection.\arabic{equation}}

\begin{center}
\huge{\textbf{Online Appendix}}
\end{center}
\vspace{0.25 cm}

\section{Quadratization Methods}
\label{sec:quadratization_methods}

We discuss four quadratization methods that introduce either no auxiliary variables or the minimum number of auxiliary variables.

\paragraph{Deduction Reduction.} \citet{TOD15} propose a quadratization strategy involving deducing properties that must hold in the ground state and performing variable substitutions accordingly. They demonstrate this on the simple constraint satisfaction problem:
\begin{gather}
    x_{1} + x_{2} + x_{3} = 1
\label{eqn:TOD15_1}\\
    x_{1}x_{4} + x_{2}x_{5} = x_{3}
\nonumber\\
    x_{1} + 2x_{2} = x_{3} + 2x_{4}.
\nonumber
\end{gather}
The problem can be expressed in PBF form as:
\begin{equation}
    \begin{split}
        \mathcal{H} &= (x_{1} + x_{2} + x_{3} - 1)^{2} + (x_{1}x_{4} + x_{2}x_{5} - x_{3})^{2} + (x_{1} + 2x_{2} - x_{3} - 2x_{4})^{2} \\
        &= 2x_{1}x_{2}x_{4}x_{5} - 2x_{1}x_{3}x_{5} - 2 x_{2}x_{3}x_{5} -2x_{2}x_{3} + 6x_{1}x_{2} - 3x_{1}x_{4} - 8x_{2}x_{4} + \\ & + x_{2}x_{5} + 3x_{2} + 4x_{3}x_{4} + x_{3}+4x_{4}+1.
    \end{split}
\label{eqn:TOD15_hamiltonian}
\end{equation}

For our purposes, it suffices to consider the first deduction-reduction. Equation \eqref{eqn:TOD15_1} suggests that $x_{1}x_{2} = x_{2}x_{3} = x_{3}x_{1} = 0$ in the ground state. Naive substitution of this deduction into Equation \eqref{eqn:TOD15_hamiltonian} reduces the Hamiltonian to the quadratic form:
\begin{equation*}
    \mathcal{H} = -3x_{1}x_{4} - 8x_{2}x_{4} + x_{2}x_{5} +
    3x_{2} + 4x_{3}x_{4} +
    x_{3} + 4x_{4} + 1.
\label{eqn:TOD15_substitution}
\end{equation*}
Whereas the original Hamiltonian had a ground state energy level of $\mathcal{H} = 0$, the new Hamiltonian has states that yield $\mathcal{H} = -3$ and $\mathcal{H} = -2$. Thus, the substitution was not valid.

\citet{TOD15} explain that this problem can be overcome by performing substitutions term-by-term and adding penalty terms where necessary. For instance, consider the quartic monomial term $2x_{1}x_{2}x_{4}x_{5}$. It can either add 0 or 2 to the Hamiltonian. In the ground state, $x_{1}x_{2} = 0$, so it will add 0. However, it may add 2 or 0 outside of the ground state. Thus, imposing $x_{1}x_{2} = 0$ will lower the energy level for some non-ground states. Thus, we must add $2x_{1}x_{2}$ as a penalty to $\mathcal{H}$ if we perform this substitution. That is, $2x_{1}x_{2}x_{4}x_{5} \rightarrow 2x_{1}x_{2}$.

\paragraph{Excludable Local Configurations (ELCs).} \citet{Ish14} defines an ELC as a partial assignment of variables that makes it impossible to achieve the ground state (global minimum). ELCs can be used to perform degree reduction without adding auxiliary variables. \citet{Dat19} provides an example of a Hamiltonian that contains an ELC: $\mathcal{H} = x_{1}x_{2} + x_{2}x_{3} + x_{3}x_{4} - 4x_{1}x_{2}x_{3}$.

Any state where $x_{1}x_{2}x_{3} = 1$ will have a lower energy state than any state where $x_{1}x_{2}x_{3} = 0$. So, the partial assignment $(x_{1},x_{2},x_{3}) = (1, 0, 0)$ is excludable, and we can apply the algorithm below, paraphrased from \citet{Ish14}, to reduce the cubic term:

\begin{enumerate}
    \item If the coefficient on the higher-order term $\zeta$ is negative, find an ELC with a parity that matches the size of the partial assignment. If it is even, then find an ELC with the opposite parity.
    \item For the ELC $a$, add the term to the Hamiltonian $\psi(x) = |\zeta| \prod_{i} \{a_{i}x_{i} + (1-a_{i})(1-x_{i}) \}$.
\end{enumerate}  

In our case, $\zeta = -4$ and the partial assignment is $(x_{1},x_{2},x_{3}) = (1, 0, 0)$. This has an odd number of variables and an odd parity (number of 1s). Thus, condition 1 is satisfied. We can compute $\psi(x)$ as in:
\begin{equation*}
    \psi(x) = |-4| (1*x_{1} + 0*(1-x_{1}))(0*x_{2} + 1*(1-x_{2}))(0*x_{3}+1*(1-x_{3})).
\label{eqn:ELC_term_example}
\end{equation*}
Adding $\psi(x)$ to $\mathcal{H}$ yields $ \mathcal{H'} = x_{1}x_{2} + x_{2}x_{3} + x_{3}x_{4} + 4x_{1} - 4x_{1}x_{2} - 4x_{1}x_{3}$, which has the same ground state as $\mathcal{H}$, but without the higher order terms.

\paragraph{Negative Term Reduction (NTR).} When a higher-order term is negative, we can use the NTR approach \citep{KZ04,FD05} to reduce it to a sum of quadratic terms using only one auxiliary variable, $x_{a}$:
\begin{equation*}
    -\prod_{i=1}^{d} x_{i} \rightarrow (d-1)x_{a} - \sum_{i=1}^{d} x_{i}x_{a}.
\label{eqn:ntr_kzfd}
\end{equation*}
\noindent This form reproduces the full energy spectrum, including the ground state. It can also reduce $d$-order terms to quadratic terms using a single auxiliary variable.

\paragraph{Positive Term Reduction (PTR).} For positive monomial terms, no algorithms can reduce $d$-order terms to quadratics using only one auxiliary variable. However, \citet{BG14} demonstrate how to reduce a $d$-order term to a quadratic using $d-2$ auxiliary variables:
\begin{equation*}
    \prod_{i=1}^{d} x_{i} \rightarrow \left(\sum_{i=1}^{d-2}x_{a_{i}}(d-i-1+x_{i}-\sum_{j=i+1}^{d} x_{j}) \right)+x_{d-1}x_{d}.
\label{eqn:ptr_bg}
\end{equation*}

\section{Annealing Schedule Simulations}\label{sec:annealing_schedule_simulations}

We consider two trivial problems that require an iterative solution. The first is encoded in the Hamiltonian $\mathcal{H}_{s} = z_{0} - z_{1} - 2 z_{0} z_{1}$. We set the initial state to $z_{0} = z_{1} = 0$ and apply Algorithm \ref{alg:oneshot_quantum} with $\mathcal{C} = 1$. If $z_{0}$ is annealed first, it will be optimal to leave its value unchanged at 0, yielding $\mathcal{H}_{s} = 0$. And if $z_{1}$ is annealed next with $z_{0}=0$, it will be optimal to flip $z_{1}$ to 1, lowering $\mathcal{H}_{s}$ to -1. In the absence of noise, an annealer would return this as the candidate solution, even though the global minimum is $\mathcal{H}_{s} = -3$ at $z_{0} = z_{1} = 1$. We must perform another iteration ($\mathcal{C} = 2$) to correct this. As shown in Table \ref{tab:pop_results}, the $\mathcal{C}=1$ case yields the correct answer in 10\% of anneals, whereas $\mathcal{C}=2$ boosts this success rate to 80\%.

\begin{table}[!ht]
\vspace{4mm}
\setlength{\tabcolsep}{5pt}
\begin{center}
\scalebox{1.0}{
\begin{tabular}{|c|c|c|}
\hline
Cycles & Share Correct ($\mathcal{H}_{s}$) & Share Correct ($\mathcal{H}_{c}$) \\
\hline 
1 & 0.10 & 0.00 \\
2 & 0.80 & 1.00 \\
\hline
\end{tabular}}
\caption{Share of anneals that yield the correct solutions in the problems $\mathcal{H}_{s}$ and $\mathcal{H}_{c}$ given how many times the group 1 and group 2 annealing subschedules are repeated within a single anneal.}
\label{tab:pop_results}
\end{center}
\end{table}

The second problem, which is more challenging and is encoded in the Hamiltonian $\mathcal{H}_{c} = z_{2}(2+z_{0}-2z_{0}z_{1}) + z_{3}(2-z_{1}-2z_{0}z_{1})$, activates and de-activates components of the loss function using auxiliary qubits, $z_{2}$ and $z_{3}$. This provides additional control over the implementation of cycles and aligns more closely with Algorithm \ref{alg:oneshot_quantum}. As with the previous problem, initializing in state $z_{0} = z_{1} = z_{2} = z_{3} = 0$ ensures that multiple cycles are needed to reliably obtain the global minimum. We show this experimentally in Table \ref{tab:pop_results}, where $\mathcal{C}=1$ yields the correct solution 0\% of the time, but $\mathcal{C}=2$ increases the success rate to 100\%.

\end{document}